\documentclass[aps,prb,showpacs,twocolumn,citeautoscript,superscriptaddress,footinbib,bibnotes]{revtex4-1}
\usepackage[dvips]{graphicx}
\usepackage{dcolumn}
\usepackage{bm}
\usepackage{fancyhdr}
\usepackage{placeins}
\usepackage{ifthen}
\usepackage{eurosym}
\usepackage{float}
\usepackage{longtable}
\usepackage{hyperref}
\usepackage{footnote}
\usepackage{rotating}
\usepackage{mathtools}
\usepackage{amsmath}
\usepackage{amssymb}
\usepackage[usenames,dvipsnames]{color}

\newcommand{\bmr}[1]{\mbox{{\bf #1}}}
\newcommand{\bfeps}{\mbox{\boldmath$\epsilon$}}

\begin{document}
\preprint{PrNiO3}
\title{Distortion mode anomalies in bulk PrNiO$_3$:  illustrating the potential of symmetry-adapted distortion mode analysis for the study of phase transitions}

\vspace{-7mm}

\author{D. J. Gawryluk}
\email{dariusz.gawryluk@psi.ch}
\altaffiliation{On leave from Institute of Physics, Polish Academy of Sciences, Aleja Lotnikow 32/46, PL-02-668 Warsaw, Poland}
\author{Y. M. Klein}
\author{T. Shang}
\affiliation{Laboratory for Multiscale Materials Experiments, Paul Scherrer Institut, 5232 Villigen PSI, Switzerland}
\author{D. Sheptyakov}
\author{L. Keller}
\affiliation{Laboratory for Neutron Scattering and Imaging, Paul Scherrer Institut, 5232 Villigen PSI, Switzerland}
\author{N. Casati}
\affiliation{Swiss Light Source, Paul Scherrer Institut, 5232 Villigen PSI, Switzerland}
\author{Ph. Lacorre}
\affiliation{Institut des Mol\'{e}cules et Mat\'{e}riaux du Mans (IMMM) - UMR 6283 CNRS, Le Mans Universit\'{e}, Avenue Olivier Messiaen, 72085 Le Mans, France}
\author{M. T.~Fern\'{a}ndez-D\'{\i}az}
\author{J.~Rodr\'{\i}guez-Carvajal}
\affiliation{Institut Laue Langevin, 71 avenue des Martyrs, CS 20156 -38042 Grenoble CEDEX 9, France}
\author{M. Medarde}
\affiliation{Laboratory for Multiscale Materials Experiments, Paul Scherrer Institut, 5232 Villigen PSI, Switzerland}
\email{marisa.medarde@psi.ch}

\date{\today}

\begin{abstract}
The origin of the metal-to-insulator transition (MIT) in $R$NiO$_3$ perovskites with \textit{R} = trivalent 4$f$ ion has challenged the condensed matter research community for almost three decades.  A drawback for progress in this direction has been the lack of studies combining physical properties and accurate structural data covering the full nickelate phase diagram. Here we focus on a small region close to the itinerant limit ($R$ = Pr, 1.5 K < $T$ < 300 K), where we investigate the gap opening and the simultaneous emergence of charge order in PrNiO$_3$. We combine electric resistivity, magnetization, and heat capacity measurements with high resolution neutron and synchrotron x-ray powder diffraction data that, in contrast to previous studies, we analyze in terms of symmetry-adapted distortion modes. Such analysis allow us to identify the contribution of the different modes to the global distortion in a broad temperature range. Moreover, it shows that the structural changes at the MIT, traditionally described in terms of the evolution of the interatomic distances and angles, appear as abrupt increases of $all$ nonzero mode amplitudes at $T_{MIT}$ = $T_{N}$ $\sim$ 130 K accompanied by the appearance of new modes below this temperature. A further interesting observation is the existence of a nearly perfect linear correlation between the amplitude of the breathing mode associated to the charge order and the staggered magnetization below the MIT. Our data also uncover a previously unnoticed anomaly at $T^* \sim$ 60 K ($\sim 0.4 \times\ T_\mathrm{MIT}$), clearly visible in the electrical resistivity, lattice parameters and some mode amplitudes. Since phase coexistence is only observed in a small temperature region around $T_\mathrm{MIT}$ ($\sim$  $\pm$ 10 K), these observations suggest the existence of a hidden symmetry in the insulating phase. We discuss some possible origins, among them the theoretically predicted existence of polar distortions induced by the non-centrosymmetric magnetic order [\href{https://doi.org/10.1088/0953-8984/28/28/286001}{J. Perez-Mato et al., J. Phys.: Condens. Matter \textbf{28}, 286001 (2016)}; \href{https://link.aps.org/doi/10.1103/PhysRevLett.103.156401}{G. Giovanetti et al., Phys. Rev. Lett. \textbf{103}, 156401 (2016)}].
\end{abstract}


\maketitle

\section{Introduction}
\begin{figure*}[tbh]
\includegraphics[keepaspectratio=true,width=165 mm]{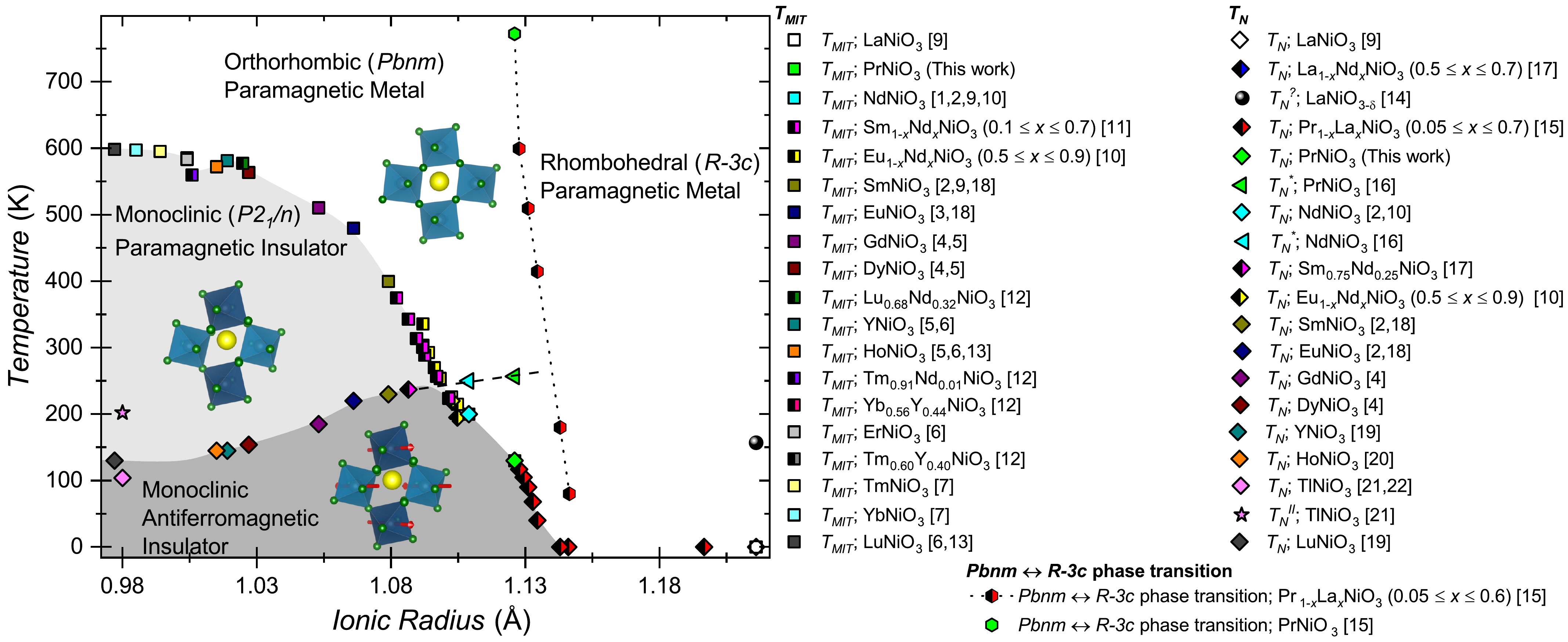}
\vspace{-3mm}
\caption{State of the art phase diagram of bulk RNiO$_3$ perovskites. The colors of the different markers indicate the reference at the origin of the data, and their shape identify the different transition temperatures (Squares: $T_{MIT}$;  Rhombuses: $T_N$ or $T_{MIT}$ = $T_N$; Triangles: linearly extrapolated $T_N$ for $R$ = Pr and Nd. Circles: the $Pbnm$ $\rightarrow$ $R\bar{3}c$ structural phase transition). For nickelates with monoclinic $P2_{1}/n$ symmetry at base temperature, the ionic radii in the abscissa correspond to the values tabulated in ref.~\cite{Shannon1976} for trivalent 4f cations in 8-coordination. For the nickelates with rhombohedral $R\bar{3}c$ symmetry at base temperature, we use instead the ionic radii for trivalent 4f cations in 9-coordination. For solid solutions, weighted averages of the trivalent 4f cations ionic radii in the coordination corresponding to the space group of the mixed-$R$ nickelate at base temperature are used.~\cite{Garcia-Munoz1992,Lacorre1991,Torrance1992,Escote2006,Frand1995,Alonso1995Eu,Alonso1999GdDy,Alonso1999SmEuGdDyHoY,Martinez-Lope2014,Alonso2001HoYErLu,Alonso1999SmEuGdDyHoY,Alonso2000,Alonso2013TmYb,Guo2018,Rosenkranz_PhD_Thesis,Vobornik1999,Frandsen2016,Rodriguez-Carvajal1998,DEMAZEAU1971,Fernandez-Diaz2001,Korosec2017,Kim2002} }
\label{fig:Phase_diagram}
\end{figure*}

The evolution of the physical properties at the crossover from localized to itinerant behavior in highly-correlated electron systems remains a fundamental problem in both, solid-state physics and chemistry.  $R$NiO$_3$ perovskites with \textit{R} = trivalent 4$f$ ion constitute a particularly well suited system to investigate this region because, in contrast to most oxide systems, a complete evolution between these two extremes can be achieved without doping. As shown in refs.~\cite{Lacorre1991,Torrance1992,Alonso1995Eu,Alonso1999GdDy,Alonso1999SmEuGdDyHoY,Alonso2001HoYErLu,Alonso2013TmYb} and illustrated in Fig.~\ref{fig:Phase_diagram},  bulk nickelates with $R$ = Pr$^{3+}$ to Lu$^{3+}$ display spontaneous metal to insulator transitions (MITs) at a temperatures $T_{MIT}$ between 130 to 600 K, allowing a continuous study of such fascinating crossover region in a particularly clean way.

In spite of intense research activity during the last 30 years, the origin of the spontaneous electronic localization observed below $T_{MIT}$ is still the subject of a lively debate~\cite{Mazin2007,Lee2011prl,Gou2011,Lau2013,Subedi2015,Mercy2017,Hampel2017,Varignon2017,Hampel2019,Peil2019}. In particular, it is not clear what is the role of the lattice, magnetic and electronic degrees of freedom, and how their involvement changes along the series. The implication of the lattice in the MIT was pointed out already in 1991, after discovering the existence of abrupt volume anomalies at the transition~\cite{Lacorre1991,Garcia-Munoz1992,Alonso2001HoYErLu,Rodriguez-Carvajal1998}. Further support came from the observation of a large $^{16}$O-$^{18}$O isotope effect on $T_{MIT}$~\cite{Medarde1998IEPrNdSmEu}, and more recently, from the possibility to control the MIT in thin films by exciting some particular vibrational phonon modes of the substrate~\cite{Caviglia2012}. The electronically unstable Jahn-Teller (JT) configuration of Ni, which is formally trivalent in these compounds ($t_{2g}^6$ $e_{g}^1$), was believed to be at the origin of these observations~\cite{Medarde1997Rev}. Interestingly, the structural modifications observed below $T_{MIT}$ are not compatible with conventional JT distortions but rather with the splitting of the unique Ni$^{3+}$ site present in the metallic state into two Ni sites with different average Ni-O distances~\cite{Alonso1999Y}. This has been interpreted as evidence of either, a 2Ni$^{3+} \rightleftarrows$ Ni$^{3+\delta}$ + Ni$^{3-\delta}$ charge disproportionation (CD),~\cite{Alonso1999Y,Medarde2009} or a Ni-O bond disproportionation with constant charge (2+) at the Ni sites and distinct amounts of holes at the O sites~\cite{Zhou2004,Johnston2014,Green2016,Bisogni2016}. These observations have led to the suggestion that the symmetry lowering and concomitant charge ordering observed in the low temperature insulating phase could be the result of the softening and further condensation of an oxygen-breathing phonon mode ~\cite{Mercy2017}, which would couple to the Ni electronic instability to open a gap below $T_{MIT}$ ~\cite{Peil2019} .

The role of the magnetism is presently less clear, on one side because the MIT exists independently of the proximity of the antiferromagnetic state observed below $T_N$ $\leq$ $T_{MIT}$, but also because of four recent reports suggesting the existence of new magnetic phases in TlNiO$_3$~\cite{Korosec2017} and LaNiO$_3$~\cite{Zhang2017,Guo2018,Wang2018}. Moreover, there is increasing evidence suggesting that the gap does not open in the same way when $T_{MIT}$ $>$ $T_N$ (Lu to Sm) than when the two transition temperatures coincide (Nd and Pr). This was first noticed in the early photoemission studies, where the sharp disappearance of the density of states (DOS) in the vicinity of the Fermi level ($E_F$) for the heavier nickelates contrasted with the persistence of non-zero DOS for PrNiO$_3$ and NdNiO$_3$ down to $T \sim 0.4 \times \, T_{MIT}$~\cite{Vobornik1999}. Further experimental evidence in this sense was suggested by the hysteretic behavior of the electrical resistivity, optical conductivity and differential scanning calorimetry measurements across the MIT, usually much larger when $T_{MIT}$ coincides with the N\'eel temperature~\cite{Liu2010,Boris2011,Liu2012,Liu2013,Hepting2014,Catalano2014,Catalano2015,Meyers2016,Middey2018}. These observations have led to the proposition that the MIT is first order when $T_{MIT}$ $=$ $T_N$, and only weakly first (or even second) order for the nickelates with $T_{MIT}$ $\neq$ $T_N$. ~\cite{Ruppen2017,Hampel2017,Peil2019}

The aim of this study is to shed new light on the order parameters involved in the MIT when $T_{MIT}$ and $T_N$  coincide. For this purpose we present here a detailed re-investigation of the crystal structure of  bulk PrNiO$_3$ across $T_{MIT}$ =  $T_{N}$ $\sim$ 130 K and within the insulating phase, that we compare with the evolution of several physical properties measured on the same sample. In contrast to early studies, we use here the correct space group (SG) in the insulating phase ($P2_{1}/n$), as well as the symmetry-adapted distortion mode formalism to analyze the neutron powder diffraction data~\cite{AMPLIMODES1,AMPLIMODES2,BilbaoCS1,BilbaoCS2,BilbaoCS3}. This approach has the advantage of enabling the decomposition of the global distortion of the crystal structure into contributions from the different modes with symmetries given by the irreducible representation of the undistorted parent SG. The evolution of the individual mode amplitudes across the transition can thus be determined. Since such amplitudes can be seen as order paramers of the different individual distortions, they allow us to describe the structural changes at the MIT and their link with the physical properties from a fully different perspective.

Using this formalism, we show here that the lattice anomalies at the MIT, traditionally described in terms of the evolution of the interatomic distances and angles, appear as sharp rises of the orthorhombic mode amplitudes accompanied by the appearance of new modes below $T_{MIT}$ = $T_N$. Our data also allow us to classify them according to their magnitude, to identify primary-secondary coupling schemas, and to establish the existence of a nearly perfect linear correlation between the amplitude of the breathing mode associated to the charge order and the staggered magnetization below $T_{MIT}$. Moreover, they uncover a previously unnoticed transient regime characterized by anomalous temperature dependence of the lattice parameters and several distortion mode amplitudes that persists down to $T^*$  $\sim$ 0.45 x $T_{MIT}$. Based on these observations, we suggest that the transient region could be the signature of a hidden symmetry in the insulating phase. We discuss some possible origins, among them the theoretically predicted existence of polar distortions induced by the non-centrosymmetric magnetic order.

\section{Experimental details}
\subsection{Sample preparation}
The PrNiO$_3$ sample (S2) used in this work was synthesized using a  high-pressure setup~\cite{Karpinski1992,Kaldis1992,Karpinski2012} recently relocated at the Paul Scherrer Institut (PSI) in Villigen, Switzerland. The starting materials were Pr$_6$O$_{11}$ (99.996\%, Sigma-Aldrich), NiO (99.99\%, Sigma-Aldrich) and aqueous (65\%) HNO$_3$ (99.999\%,  Alfa Aesar). Pr$_6$O$_{11}$ and NiO were dried in air at 750 $^\circ$C and 300 $^\circ$C respectively for 24 h, and then cooled to 150 $^\circ$C. Freshly dried Pr$_6$O$_{11}$ (5.1043 g, 4.997 mmol) and NiO (2.2395 g, 29.984 mmol) were suspended in deionized water (20 ml), and aqueous HNO$_3$  (80 ml) was added in small portions. The clear green solution was heated at 350 $^\circ$C for $\sim$ 4 - 6 h to evaporate water and decompose the nitrates. The nitrates were further decomposed by heating in air in a furnace at 350 $^\circ$C for 6h with intermediate regrinding into a fine powder. The black powder was annealed at 650 $^\circ$C under 1 bar of oxygen flow for 48 h and cold pressed into a 5 mm rod using a hydrostatic press by applying 4000 bar uniaxial pressure. The rod was once annealed at 1150 $^\circ$C and 167 bar oxygen pressure (c.a. 14 cm$^3$ in volume) in a high pressure furnace for 24 h to yield perovskite PrNiO$_3$ (5.4353 g, 21.952 mmol, 73.2\% based on Ni). Laboratory x-ray powder diffraction indicated that the sample was very well crystallized and free of impurities. Part of the as-prepared rod was cut in small pieces (15-30 mg) for physical property measurements. The rest was grinded into a fine powder that was subsequently used for x-ray and neutron  diffraction measurements.

\begin{figure}[tbh]
\includegraphics[keepaspectratio=true,width=\columnwidth]{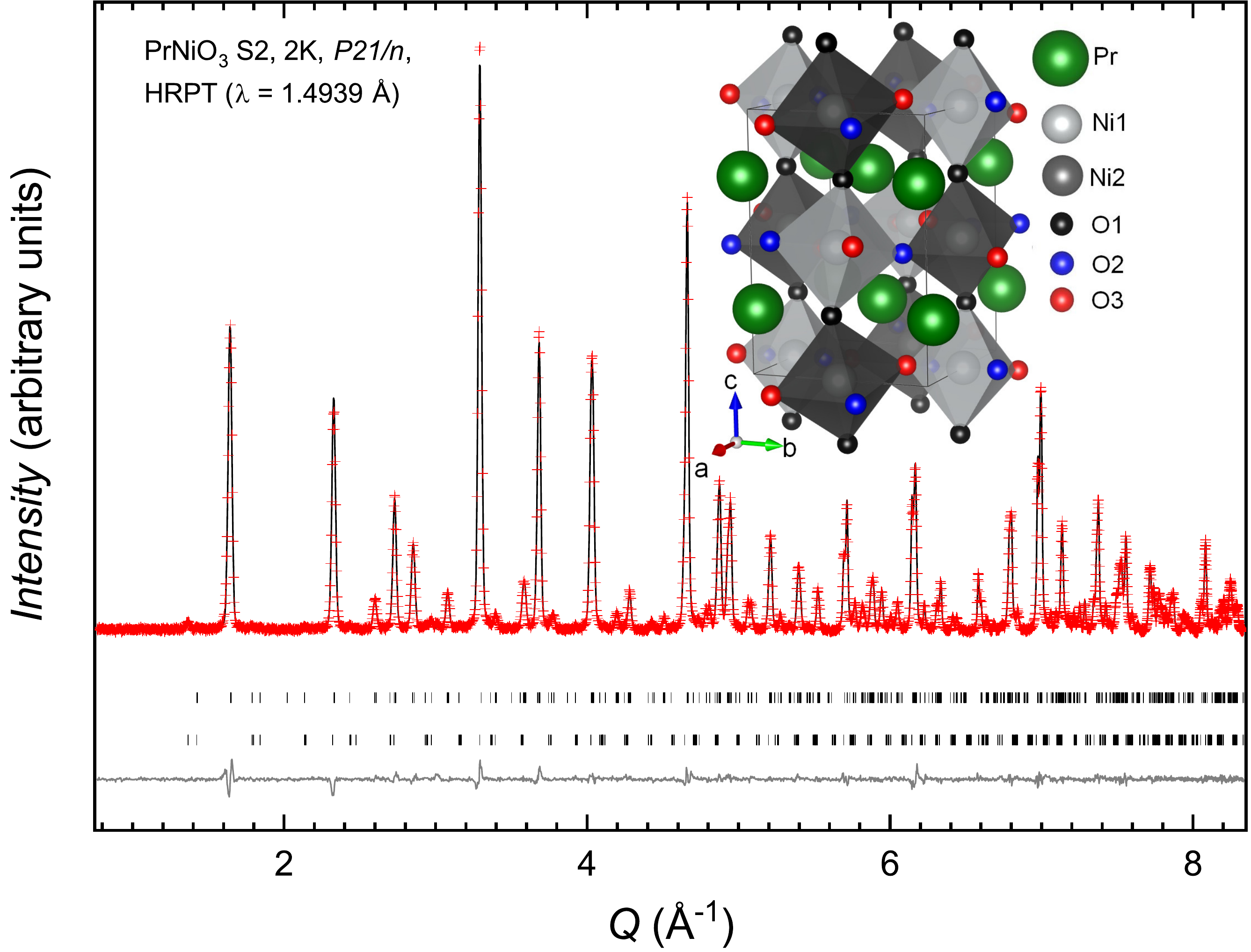}
\vspace{-3mm}
\caption{ Rietveld fit of the neutron powder diffraction patterns  for PrNiO$_3$ (S2) measured at 2 K in HRPT. Red crosses: observed data. Black lines: calculated and difference patterns. The vertical ticks indicate the positions of the Bragg reflections for the crystal structure (upper row) and the collinear antiferromagnetic magnetic structure with \emph{\textbf{k}} = (1/2, 0 ,1/2) (lower row) described in the text. Inset:  low temperature ($T$ $<$ $T_{MIT}$) crystal structure of PrNiO$_{3}$.}
\label{fig:HRPT_2K}
\end{figure}

\subsection{Neutron and synchrotron x-ray diffraction}
The synchrotron x-ray  powder diffraction data were obtained at the Material Science Beam Line (X04SA) of the Swiss Ligh Source (SLS) of the PSI~\cite{Willmott2013}.  The sample was loaded in a borosilicate glass capillary (D = 0.1 mm, $\mu$R = 0.29)  and measured in transmission mode with a rotational speed of $\sim$ 1 Hz and  Si (111)  $\lambda$ = 0.49226 \AA \  (\textit{Q}$_{max}$ =  18 {\AA}$^{-1}$). The primary beam was vertically focused and slitted to about 300 x 4'000 ${\mu}$m$^2$. Powder diffraction patterns were recorded for 10 s at temperatures between 80 and 500 K using an Oxford Nitrogen cryojet and a Mythen II 1D multistrip detector with energy discrimination (2$\theta_{max}$ = 90$^\circ$, 2$\theta_{step}$ = 0.0018$^\circ$), and then binned into one pattern. The wavelength, zero offset and resolution function were determined using a LaB$_6$ powder standard (NIST SRM 660c). The measurements were performed by heating after a stabilization time of about 3 minutes with a typical acquisition times of 15 minutes per temperature.

The neutron powder diffraction (NPD) measurements were carried out at the Swiss Spallation Neutron Source (SINQ) of the PSI. About 5g of PrNiO$_3$ powder were introduced into a cylindrical vanadium can (D = 6 mm, H = 5 cm)  and mounted on the stick of a He cryostat, whose contribution to the powder diffraction patterns was eliminated using an oscillating radial collimator.  Several patterns were collected at  selected temperatures between 2 and 205 K at the high resolution diffractometer HRPT ~\cite{Fischer2000}  (primary beam collimations $\alpha_1$ = 12';  Ge(511)  $\lambda$ = 1.494 \AA \ (\textit{Q}$_{max}$ = 8.33 {\AA}$^{-1}$);  2$\theta_{step}$ = 0.05$^\circ$). All measurements were performed by heating  after a stabilization time of 15 minutes with typical acquisition times of 3 hours per temperature. The exact wavelength was refined from a combined refinements of the HRPT neutron and the X04SA synchrotron x-ray diffraction data at 145K.  A second set of patterns was collected on the cold neutron diffractometer DMC~\cite{Schefer1990} (pyrolytic graphite (002); $\lambda$ = 2.458 \AA \;   2$\theta_{step}$ = 0.1$^\circ$) while ramping  the temperature from 2 to 300 K at 0.2 K/minute. A  further pattern with high statistics was recorded at 2 K. The exact value of the wavelength was adjusted to match the lattice parameters refined at high resolution instruments at the same temperature.

\begin{figure}[tbh]
\includegraphics[keepaspectratio=true,width=\columnwidth]{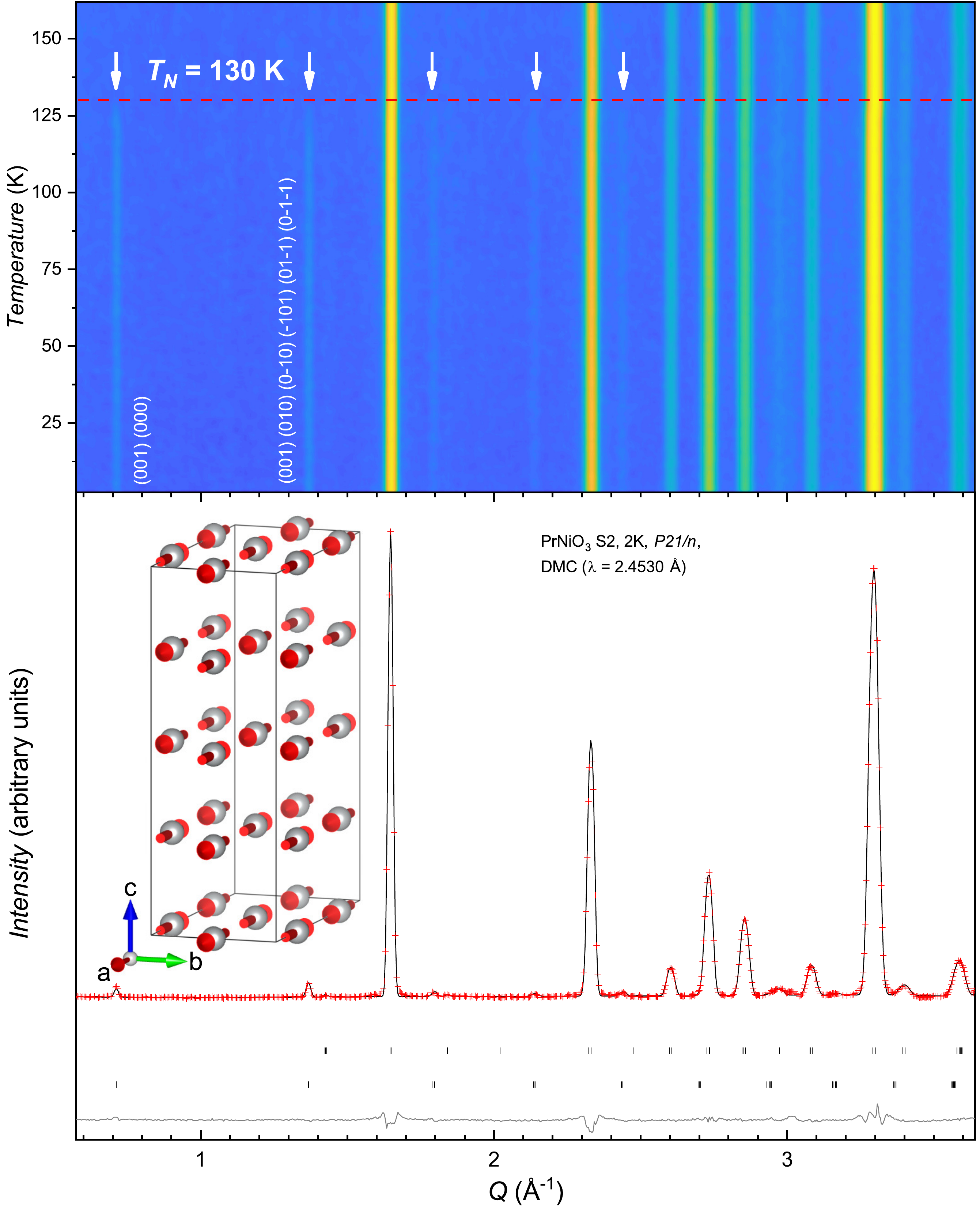}
\vspace{-3mm}
\caption{ Upper panel: 2D contour plot showing the temperature dependence of the NPD patterns for PrNiO$_3$ (S2) measured on DMC. The white arrows indicate the position of the magnetic reflections. Lower panel: Rietveld fit of the neutron powder diffraction patterns measured at 2 K on DMC. Red crosses: observed data. Black lines: calculated and difference patterns. The vertical ticks indicate the positions of the Bragg reflections for the crystal structure (upper row) and the collinear antiferromagnetic magnetic structure with \emph{\textbf{k}} = (1/2, 0, 1/2) (lower row) described in the text. Inset:  collinear magnetic structure of PrNiO$_{3}$ used in the fit (see text for details). }
\label{fig:DMC_2K}
\end{figure}

All powder diffraction data were analyzed with the Rietveld package FullProf Suite~\cite{FullProf1}. For the refinements of the NPD data, more sensible to the oxygen positions, we employed the symmetry-adapted distortion mode formalism for the description of the crystal structure. The mode decomposition was performed using the program $\textit{AMPLIMODES}$~\cite{AMPLIMODES1,AMPLIMODES2}, freely available at the Bilbao Crystallographic Server~\cite{BilbaoCS1,BilbaoCS2,BilbaoCS3}. Below $T_{MIT}$  = $T_N$ new reflections corresponding to the \emph{\textbf{k}} = (1/2, 0, 1/2) magnetic propagation vector were observed. To describe them, a second phase was added to the fits. Representative patterns of the Rietveld refinements of  HRPT and DMC data, as well as a contour plot showing the evolution of the NPD patterns recorded on DMC between 2 and 162 K are shown in Figs.~\ref{fig:HRPT_2K} and ~\ref{fig:DMC_2K}.

\subsection{Physical properties}
DC magnetization measurements  were performed in a superconducting quantum interference device magnetometer (MPMS-XL 7T, Quantum Design) between 2 and 400 K while heating under a magnetic field of $\mu_0H$ = 0.5 T after being cooled down to 2 K in zero field.  Heat capacity measurements between 2 and 300 K were carried out on Physical Property Measurement System (PPMS 9T, Quantum Design) by heating using the relaxation method. The same instrument was used to measure the electrical resistivity between 2 and 300 K (both on cooling and heating) using the four-point method. The sample was contacted with Pt wires fixed with silver paste.

\section{Results}
\subsection{Anomalies in the physical properties }
\subsubsection{Electrical resistivity }
Fig.~\ref{fig:Physical_properties}a shows the temperature  ($T$) dependence of the electrical resistivity ($R$) measured in a cooling-heating cycle . The sharp anomaly at $T_{MIT}$ reported in previous studies is clearly observed, both by heating and cooling. The $T_{MIT}$  value, as defined by the highest value of the first derivative, is 129 K (heating)  and 117 K (cooling). The resistivity at 2.5 K is $\sim$ 10$^6$ times larger than the value above $T_{MIT}$ and is similar to the best values reported for thin films with comparable transition temperatures \cite{Stemmer2013}. This, together with the very small difference between the $T_{MIT}$ values measured by heating and cooling (12 K)  testify to the high sample quality.

Below the transition the resistivity displays a pronounced hysteretic behavior down to  90 K, followed by a slope change at $\sim$ 55 K. This is better appreciated in  Fig.~\ref{fig:Physical_properties}b, showing the first derivative of the \textit{RT} product as a function of  $T$.  Given that $T$ is a monotonic variable, any change in its derivative reflects slope changes in $R$. We note that the heating and cooling curves in Fig.~\ref{fig:Physical_properties}b are very different between the transition and 90 K, most probably due to the coexistence of the metallic and insulating phases within this temperature range. Interestingly, both curves become identical below 90 K, indicating that below this temperature, the thermal variation of $R$ does not depend on the cooling/heating sense.

\subsubsection{Heat capacity }
The temperature dependence of the heat capacity $C_p$, not reported previously to our best knowledge, is shown in Fig.~\ref{fig:Physical_properties}c. The only noticeable anomaly of the curve, measured by heating, is an extremely sharp peak at $T_{MIT}$ =  $T_{N}$ = 129 K, in excellent agreement with the value obtained from the resistivity curve measured by heating. This results in a sharp discontinuity also in $C_p$/$T$ (Fig.~\ref{fig:Physical_properties}) and the entropy $S$. A proper evaluation of the $\Delta S$ jump at the transition, which requires estimations of the lattice contribution in absence of volume anomaly, the contribution of the Pr$^{3+}$ crystal field levels, and the electronic contribution due to the delocalization process, will require further experimental and theoretical work, and will be presented separately. The sharpness of the entropy anomaly at its non-lambda shape give nevertheles further support to the 1$^{st}$ order nature of the transition suggested by the hysteretic behavior of the resistivity between 90 K and the MIT, in full agreement with previous reports.

\subsubsection{Magnetic susceptibility }
\begin{figure}[tbh!]
\includegraphics[keepaspectratio=true,width=\columnwidth]{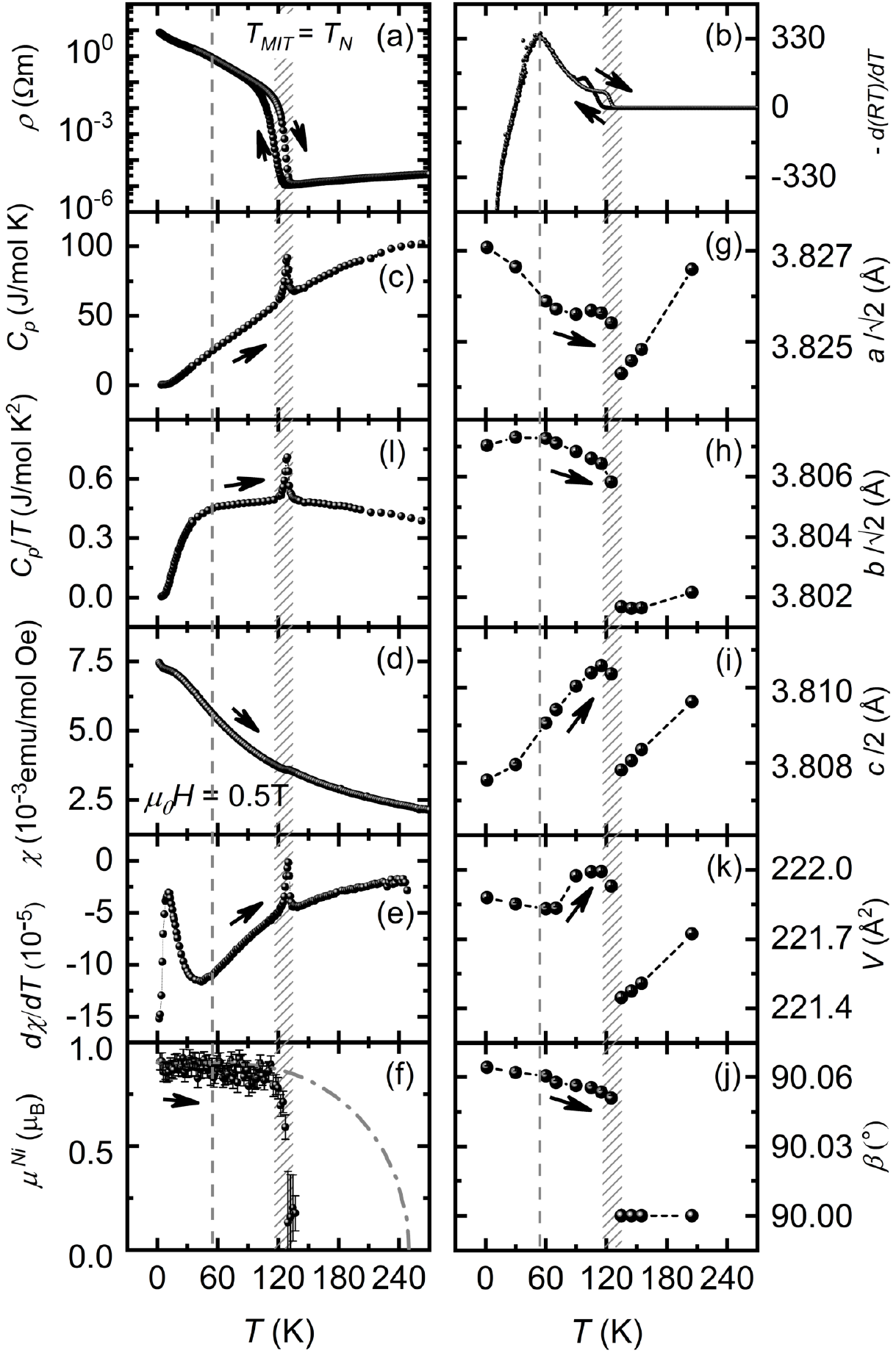}
\vspace{-3mm}
\caption{Temperature dependence of selected physical properties and lattice parameters of PrNiO$_3$ (S2). a) Electric resistivity. b) First derivative of the \textit{RT} product,  multiplied by (-1). c) Heat capacity $C_p$. d) Magnetic susceptibility $\chi$ = M/$\mu_0H$  measured in a magnetic field of 0.5 T. l) $C_p$/T. e) First derivative $d \chi /dT$ of the magnetic susceptibility. f) Staggered magnetization of Ni from neutron powder diffraction. The circles correspond of the results obtained using the collinear model with identical Ni sites described in the text. The discontinuous line corresponds to the calculation using a Brillouin function with S = 1/2 and $T_N^*$ = 257 K.   g-k) Temperature dependence of unit cell parameters $ a / \sqrt{2}$, $ b / \sqrt{2}$, $c$/2, the monoclinic angle $\beta$, and the unit cell volume $V$. Error bars are smaller than data points indicators, and the dashed lines are guides for the eye. The dashed rectangle around $T_{MIT}$ indicates the region where phase coexistence is clearly observed in the synchrotron XRD data. The vertical dashed line shows the anomaly at $\sim$ 55 K.}
\label{fig:Physical_properties}
\end{figure}

The magnetic susceptibility $\chi$ = $M/\mu_0H$  measured by heating under a magnetic field of 0.5 T is shown in  Fig.~\ref{fig:Physical_properties}d. Although the curve is dominated by the paramagnetic contribution of Pr (nominally Pr$^{3+}$), the first derivative (Fig.~\ref{fig:Physical_properties}e)  shows the presence of two clear anomalies. The first one, extremely sharp, peaks at $T_N$ = $T_{MIT}$ = 130 K, in excellent agreement with the value extracted from the resistivity measured by heating. The second, consisting of  a broad maximum peaking around $\sim$ 10 K, is most probably related to the magnetism of the praseodymium. Since no evidence of neither, polarization nor cooperative magnetic order of the Pr magnetic moments could be observed in the neutron powder diffraction patterns, the anomaly could be related with the thermal population of the Pr$^{3+}$ crystalline electric field levels~\cite{Rosenkranz1999}.

The  temperature dependence of the Ni staggered magnetization derived from neutron powder diffraction (Fig.~\ref{fig:Physical_properties}f) is again in agreement with the behavior expected for a first-order transition and consistent with previous reports~\cite{Garcia-Munoz1994,Vobornik1999}.  The value of the Ni magnetic moment, obtained from the Rietveld fits performed using a collinear magnetic structure with identical moments in the two Ni sites (see next section for details), is nearly saturated below $T_N$, reaching a value of 0.95 (3) $\mu_{B}$ at 1.5 K. This value is very close to the  single-ion, spin-only Ni$^{3+}$ moment (1 $\mu_{B}$). We note also that, according to the evolution of $T_N$ with the R ionic radius,  the  N\'eel temperature of PrNiO$_3$ in absence of  MIT is expected to be much higher ($\sim$ 257 K by extrapolating linearly as in shown in Fig.~\ref{fig:Phase_diagram}). The calculated temperature dependence of the staggered magnetization using a Brillouin function with  S = 1/2 and $T_N$ = 257 K, shown as a dashed line in Fig.~\ref{fig:Physical_properties}f, reproduces indeed very well the experimental observations below $T_N$. This is consistent with the idea that the magnetic order, fully suppressed above $T_{MIT}$, just reacts to the sudden metallization, pointing towards a dominant role of the electronic degrees of freedom in the transition. ~\cite{Subedi2015,Hampel2017,Hampel2019,Peil2019}

\subsection{Anomalies in the lattice parameters, interatomic distances and angles}

The temperature dependence of the lattice parameters of bulk PrNiO$_3$ between 1.5 K and RT has been reported in the past from the analysis of low resolution NPD data and using the orthorhombic SG $Pbnm$ for the full temperature range~\cite{Garcia-Munoz1992}.  Figs.~\ref{fig:Physical_properties}g-k show the evolution of the pseudocubic lattice parameters $a_{p} = a/\sqrt{2}$,  $b_{p} = b/\sqrt{2}$, $c_{p} = c$/2 and the monoclinic angle $\beta$ and the unit cell volume between 2 and 250 K obtained  from the high resolution NPD patterns measured on HRPT. The data were measured by heating, and fitted using the monoclinic SG $P2_1/n$ below $T_{MIT}$. The sharp anomalies at the transition reported in ref.~\cite{Garcia-Munoz1992} are clearly observed.  However, our data reveal the presence of  a further discontinuity at  $\sim$ 60 K, particularly pronounced in the case of $a_{p}$ but clearly visible in all lattice parameters. We note also that the anomaly is very close to the slope change observed in the resistivity around $\sim$ 55 K.

\begin{figure}[tbh]
\includegraphics[keepaspectratio=true,width=\columnwidth]{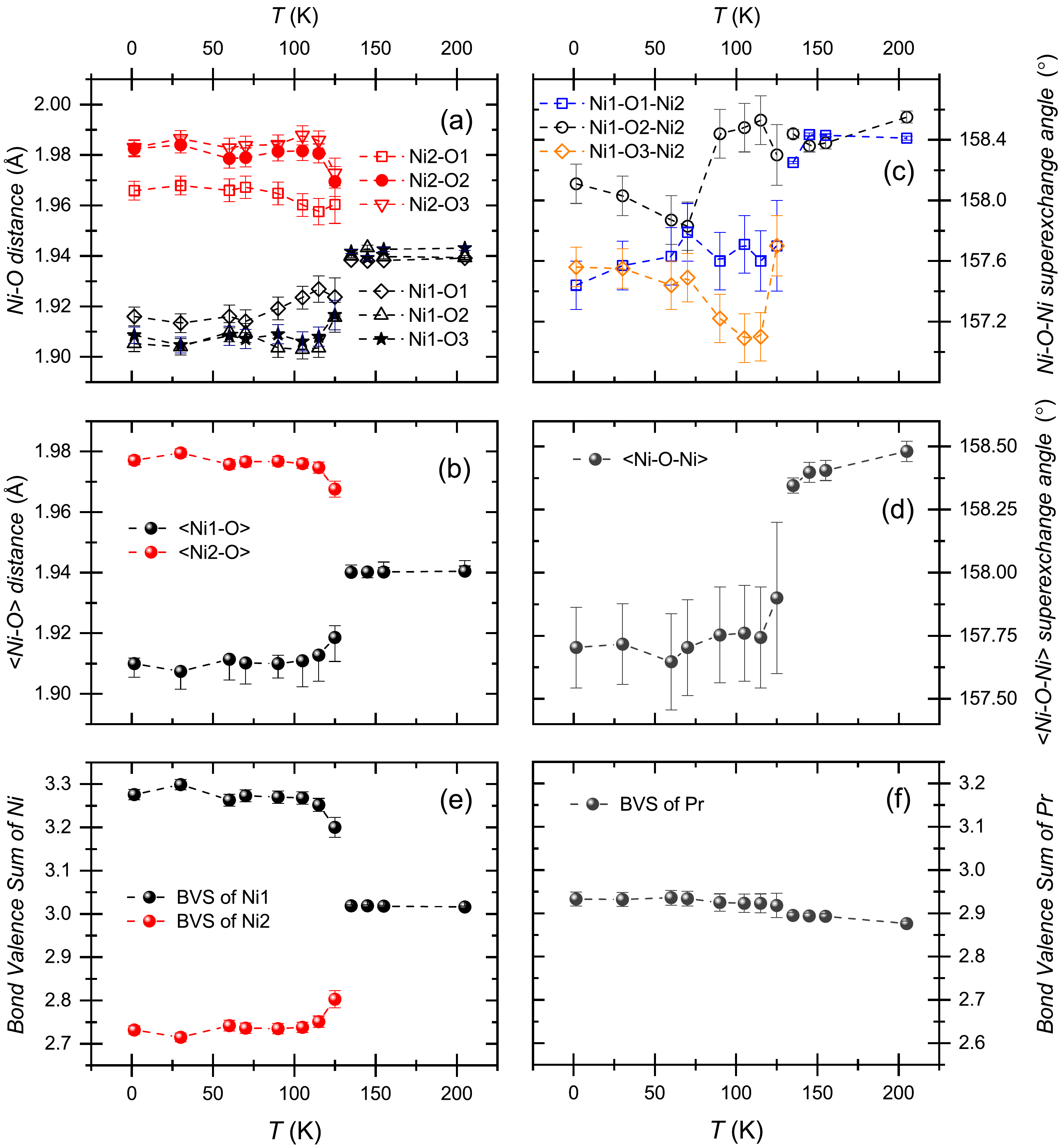}
\vspace{-3mm}
\caption{Temperature dependence of the individual and average Ni-O interatomic distances (a-b), Ni-O-Ni superexchange angles (c-d) and cations'Bond Valence Sum model (e-f) based on HRPT data for  PrNiO$_3$ (S2).}
\label{fig:Distances_angles_BVS}
\end{figure}

The anomalies of the lattice parameters at the MIT have been traditionally interpreted in terms of the changes in the Ni-O interatomic distances and Ni-O-Ni superexchange angles. In order to illustrate the full compatibility of previously reported results with those obtained using the distortion mode formalism we present in Figs.~\ref{fig:Distances_angles_BVS}a-d the evolution of the most relevant interatomic distances and angles obtained using this last method. The splitting of the single NiO$_6$ octahedron present in the metallic state into two NiO$_6$ octahedra with different average Ni-O distances at $T_{MIT}$ is clearly observed, in very good agreement with our previous work using a different PrNiO$_3$ sample~\cite{Medarde2008} (note that we use here the same atom labeling, see inset of Fig.~\ref{fig:HRPT_2K} and Supplementary Materials~\cite{Supple}). A small but clear anomaly around 60 K is also observed in all Ni-O distances. This anomaly could not be observed in ref.~\cite{Medarde2008} due to the lack of experimental points between 10 and 100 K.

The evolution of the Ni-O-Ni superexchange angles across the transition has only been reported reported using the SG $Pbnm$ in the insulating phase ~\cite{Garcia-Munoz1992}. Figs.~\ref{fig:Distances_angles_BVS}c-d show the results obtained in this study using the monoclinic SG $P2_1/n$. While the two symmetry-allowed angles are very similar in the metallic phase, this is not the case anymore below  $T_{MIT}$. The two identical in-plane angles become distinct (Ni1-O2-Ni2 and Ni1-O3-Ni2), and undergo sharp decreases at the transition. The out-of-plane angle Ni1-O1-Ni2 behaves differently, remaining nearly constant down to 90 K after what it suddenly decreases at $\sim$ 60 K. This is reflected in a small anomaly in the average superexchange angle <Ni1-O-Ni2>, which decreases abruptly below $T_{MIT}$. Moreover, it is consistent with a smaller Ni 3d - O 2p overlap (and hence narrower bandwidth) in the insulating phase, in full agreement with early reports using the SG $Pbnm$.

The Bond Valence Sums (BVS) of the Ni and Pr sites calculated using the Ni-O /Pr-O  interatomic distances and the RT BV parameters for Ni$^{3+}$, O$^{2-}$ and (8-coordinated) Pr$^{3+}$ from refs.~\cite{Altermatt1985,Brown1985} are shown in Figs.~\ref{fig:Distances_angles_BVS}e-f.  The values obtained for the two Ni sites Ni$^{3+\delta}$ and Ni$^{3-\delta}$ at 2 K  imply a $\delta$ = 0.27(2), slightly larger than in our previous work ($\delta$ $\sim$  0.21(2)), but still smaller that the $\delta$ = 0.33(2)  value reported for LuNiO$_3$ at RT. Although modest, we note that the change of the Ni BVS at $T_{MIT}$ is about 1 order of magnitude larger than the variation observed at the Pr sites, which, in absence of an electronic anomaly, is just due to the thermal contraction of the Pr-O bonds (Fig.~\ref{fig:Distances_angles_BVS}f).

\subsection{Magnetic structure and staggered magnetization}
The PrNiO$_3$ magnetic structure, characterized by the propagation vector \textbf{k} = (1/2, 0, 1/2), was refined in early studies using the SG $Pbnm$, where all Ni sites are described with the (single) Wyckoff position 4$b$ (0.5, 0, 0) ~\cite{Garcia-Munoz1992}. In this work we use instead the SG $P2_1/n$ with Ni sites splited in two distinct positions: 2$d$ (0.5, 0, 0) and 2$c$ (0.5, 0, 0.5) to fit the data obtained at 2 K on DMC. We conducted refinements with both, the collinear~\cite{Garcia-Munoz1994} and non-collinear~\cite{Fernandez-Diaz2001} models proposed in previous works while refining the moments of the two Ni sites jointly (i.e., forcing then to have the same value), and separately (letting them converge to different values).

As shown in the SUPP3 file, all models give very similar reliability indexes (slightly better for the models with different Ni moments), making impossible to disentangle them from the present data. To investigate the temperature dependence of the Ni magnetic moments we thus used the simplest of the four models (collinear with equal moments along \emph{\textbf{a}}), whose results are shown in Fig.~\ref{fig:Physical_properties}.

\subsection{Phase coexistence}
\begin{figure}[tbh]
\includegraphics[keepaspectratio=true,width=\columnwidth]{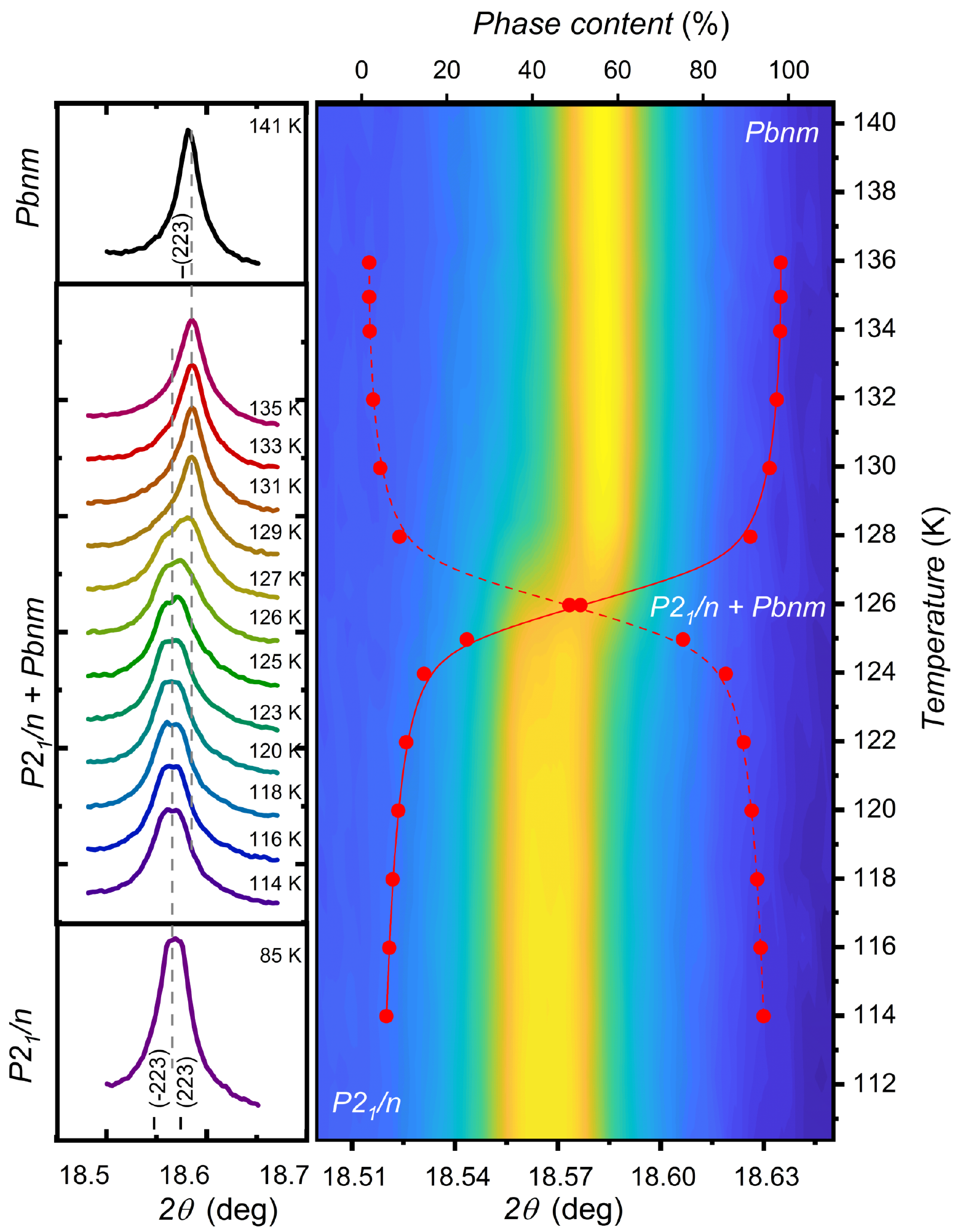}
\vspace{-3mm}
\caption{Contour plot showing the temperature dependence of the (-223)+(223) reflection in the synchrotron-x-ray powder diffraction patterns  for PrNiO$_3$ (S2) across the MIT measured by heating. The solid/dashed red lines indicate respectively the thermal evolution of the metallic ($Pbnm$) and insulating ($P2_1$/n) phase fractions.}
\label{fig:Phase_coexistence}
\end{figure}

In order to check a possible link between the anomaly observed in different structural parameters around 60 K and the coexistence of the metallic and insulating phases we followed their signature in the synchrotron x-ray diffraction patterns.  Fig.~\ref{fig:Phase_coexistence}  shows the temperature dependence of the orthorhombic reflection (223) and the monoclinic doublet (-223)+(223)  between 110 and 141 K measured by heating. Above 135 K, i.e., well above $T_{MIT}$, only the $Pbnm$ (223) reflection is present. However, a shoulder corresponding to the (-223)+(223) doublet characteristic of the $P2_1$/n insulating phase starts to be visible below this temperature. The insulating fraction, which increases slowly down to 130 K, grows much faster below this temperature, and at $\sim$ 126 K  both phases are present in nearly equal amounts.

Below 126 K the metallic fraction decreases, but the transformation rate is clearly slower than above the transition. Such asymmetry in the transformation rate leads to a significantly larger coexistence region below $T_{MIT}$. Although the temperature range of $macroscopic$ phase coexistence, i.e., with both phases present at least at 1\% level, is restricted to a very narrow temperature range around $T_{MIT}$ ($\sim$  $\pm$ 10 K), the signature of both phases in the diffraction patterns can be tracked down to almost 90 K. This is consistent with the lower phase coexistence limit inferred from electric resistivity measurements. Moreover, it suggests that, by heating, small metallic domains with sizes close to the detection limit of XRD ($<$ 4 - 5 nm) may exist between 90 and 116 K. The anomaly in the resistivity, the lattice parameters, interatomic distances and angles observed around at 55 - 60 K, also measured by heating, can be thus hardly attributed to the nucleation of the metallic phase in the insulating matrix, which most probably starts around 90 K.

\section{Distortion mode analysis}

Having shown in the previous sections that the distortion mode formalism, as implemented in the $FullProf$ $Suite$, is able to reproduce the evolution of the interatomic distances and angles reported in previous studies, we describe now the lattice anomalies at the MIT from a fully different perspective, namely, by tracking the evolution of the distortion mode amplitudes as a function of temperature. Such amplitudes can be seen as order parameters for the different individual distortions. Hence, their relative values, temperature dependence and anomalies at $T_{MIT}$ encode the information on the most relevant degrees of freedom involved in the global distortion and their evolution across the transition.

The description of a crystal structure in terms of symmetry modes involves some significant differences with respect to the usual description in terms of fractional atomic coordinates $\bmr{r}(\mu)$. An important point is that such description can only be used if the crystal structure can be considered pseudosymmetric with respect to some "parent" configuration of higher symmetry. Such parent phase may actually exist, but it can be also a virtual reference structure. A necessary condition is the existence of a group–subgroup relation between the space groups of the parent and the actual structure. In this case, the latter can be then described as the high symmetry structure plus a static, symmetry-breaking structural distortion. Such distortion can be then decomposed into contributions from different "frozen" modes (i.e., collective correlated  atomic displacements) with symmetries given by irreducible representations ($irreps$) of the parent space group~\cite{AMPLIMODES2}. Since the basis vectors or the irreps are fixed by symmetry, the only free parameters are the amplitudes of the different modes, which can be refined in the same way as the atomic coordinates in a standard least-square fit. The number of free parameters is indeed identical in both descriptions.  Here we have used AMPLIMODES for the mode decomposition, and the different mode amplitudes have been obtained from the fits of high resolution neutron powder diffraction data using FullProf Suite. To the best our knowledge, this is presently the only freely available Rietveld code allowing the use both, atomic coordinates and mode amplitudes, for the description of the crystal structure.

\subsection{Formalism}

Let's define  $\bmr{r}(\mu)$ as the vector position of the atom $\mu$ ($\mu =1...s $  ) within the asymmetric unit of the parent structure with SG $\bmr{H}$ (higher symmetry). The asymmetric unit of the observed distorted structure with lower symmetry SG $\bmr{L}$, subgroup of $\bmr{H}$, will in general have a larger number of atoms due to the splitting of the Wyckoff orbits in $\bmr{H}$. Thus, the atom positions of the structure, described in the subgroup $\bmr{L}$, are given in terms of the atom positions of the parent group $\bmr{H}$ described in the basis (unit cell) of the subgroup $\bmr{L}$ as:

\begin{equation}
{\bmr r}(\mu,i) = {\bmr r}_0(\mu,i) + {\bmr u}(\mu,i) ;\quad  \mu=1,2, ... s; i=1,2,...n_{\mu}
\end{equation}
\begin{equation}
{\bmr u}(\mu,i)=\sum\limits_{\tau,m} A_{\tau,m} \bfeps (\tau,m|\mu,i)
\end{equation}

The displacement vectors $\bmr{u}(\mu,i)$  are written as linear combinations of the polarization modes $\bfeps$ (basis vectors of the \emph{irreps} involved in the phase transition). The indices $\tau$ and $m$ label all possible distinct allowed symmetry-adapted distortion modes. The index $\tau$  stands for the different \emph{irreps}, while $m$ ($m =1,2,...n_{\tau}$) enumerates the possible different independent modes within a given \emph{irrep}, and the amplitudes $A_{\tau,m}$ encode the magnitude of the distortions associated to each mode. The normalization of the polarization vectors is such as the values of amplitudes are directly in length units of Angstroms. In order to enhance the role of the different \emph{irreps} we re-define the displacement vectors as:

\begin{equation}
{\bmr u}(\mu,i)=\sum\limits_{\tau,m} A_{\tau,m} \bfeps(\tau,m|\mu,i)= \sum\limits_{\tau} {A_\tau} {\bmr e}(\tau|\mu,i)
\end{equation}

in which the global amplitude of the \emph{irrep} ${A_\tau}$ and the new polarization vectors are defined as:

\begin{equation}
{A_\tau } = {\left( {\sum\limits_m {A_{\tau,m}^2\;} } \right)^{1/2}}
\end{equation}
\begin{equation}
{\bmr e}(\tau|\mu,i) = \sum\limits_{m} a_{\tau,m} \bfeps(\tau,m|\mu,i)
;\quad a_{\tau,m}=\frac{A_{\tau,m}} {{\left( {\sum\limits_m {A_{\tau,m}^2} } \right)}^{1/2}}\quad
\end{equation}

By examining which global amplitudes $A_{\tau }$ are nonzero we can identify the $irrep(s)$ actively contributing to the distortion of the structure.

\subsection{SGs setting and choice of the origin}

As input for $AMPLIMODES$ we used the cubic primitive perovskite lattice (SG $Pm\bar{3}m$) with one PrNiO$_3$ formula per unit cell, and the non-standard $Pbnm$ and $P2_1/n$ settings with $Z$ = 4 and \emph{\textbf{c}} as longest crystal axis for the orthorhombic and monoclinic structures. In order to ease the comparison between the parent and the distorted space groups we locate the Ni atom of the asymmetric unit at the origin. Note that this choice differs from the convention used in most experimental papers, where Ni is at (1/2 0 0). The transformation from the high symmetry into the low symmetry setting is \emph{\textbf{a}} = \emph{\textbf{a$_p$}} - \emph{\textbf{b$_p$}},  \emph{\textbf{b}} = \emph{\textbf{a$_p$}} + \emph{\textbf{b$_p$}}, \emph{\textbf{c}} = \emph{\textbf{2c$_p$}}; (0, 0, 0), the last vector indicating the origin of the low-symmetry cell expressed in the basis of the high-symmetry cell. Files with the $AMPLIMODES$ input and the obtained mode decomposition (in FullProf .pcr format) for the two space groups are provided in the Supplementary Material. In the following, \emph{\textbf{a}}, \emph{\textbf{b}} and \emph{\textbf{c}} will denote the low-symmetry lattice parameters, either of $Pbnm$ or of $P2_1/n$.

\subsection{Mode decomposition in the metallic phase}
Seven distortion modes corresponding to five different $irreps$ are allowed in the case of $Pbnm$. They are labeled $R^{+}_4$, $M^{+}_3$, $X^{+}_5$, $R^{+}_5$, and $M^{+}_2$, the capital letter indicating their \emph{\textbf{k}}-vector in the first Brillouin zone of the cubic primitive $Pm\bar{3}m$ parent cell. The modes of $R^{+}_4$ and $M^{+}_3$ symmetry involve rotations of the NiO$_6$ octahedra around the \emph{\textbf{b}} and \emph{\textbf{c}} axes, respectively, whereas $M^{+}_2$ is an in-plane stretching mode acting on the basal oxygens. These three modes involve only oxygen displacements. In contrast, those associated to the  $X^{+}_5$ and $R^{+}_5$ $irreps$ allow both, apical oxygen and rare-earth displacements.

Schematic representations of the seven \textit{Pbnm}-allowed modes are shown in Fig.~\ref{fig:Global_mode_amplitudes} together with the temperature dependence of the global amplitudes associated to each $irrep$.  The arrow lengths are arbitrary for the $irreps$ with a single mode ($R^{+}_4$, $M^{+}_3$ and $M^{+}_2$), but for those with more than one ($X^{+}_5$ and $R^{+}_5$), they reflect the refined relative amplitudes of the individual O and Pr modes. The temperature dependence of both, the global and the individual mode amplitudes for the last two $irreps$ is shown in Fig.~\ref{fig:Individual_mode_amplitudes}. A list of the basis vectors associated to each $irrep$ is provided in the SUPP1 file of the Supplemental Material.

For the individual mode amplitudes of Fig.~\ref{fig:Individual_mode_amplitudes} and the tables of the Supplemental files SUPP1 and SUPP2 we use the $AMPLIMODES$ notation \emph{AnIR}(\text{\AA}) (e.g. $A8X^{+}_5$(\text{\AA}) ), in which \emph{IR} is the symbol of the \emph{irrep} and \emph{An} (with \emph{n}=1,2,...) labels the amplitudes of the $n$ individual modes allowed by the SG.

\subsection{Mode decomposition in the insulating phase}
In the case of the monoclinic SG $P2_1/n$, five further distortion modes are allowed. Three of them involve only oxygen displacements and correspond to the new $irreps$ $R^{+}_1$ (breathing distortion), $R^{+}_3$ (Jahn-Teller distortion) and $M^{+}_5$ (basal oxygen rotation around the \emph{\textbf{a}} axis). The two remaining ones have $R^{+}_5$ symmetry and correspond respectively to praseodymium displacements along \emph{\textbf{a}}, and stretching distortion involving the basal oxygens. A total of four modes with $R^{+}_5$ symmetry are thus allowed in the monoclinic insulating phase.

\begin{figure*}[tbh!]
	\includegraphics[keepaspectratio=true,width=154 mm]{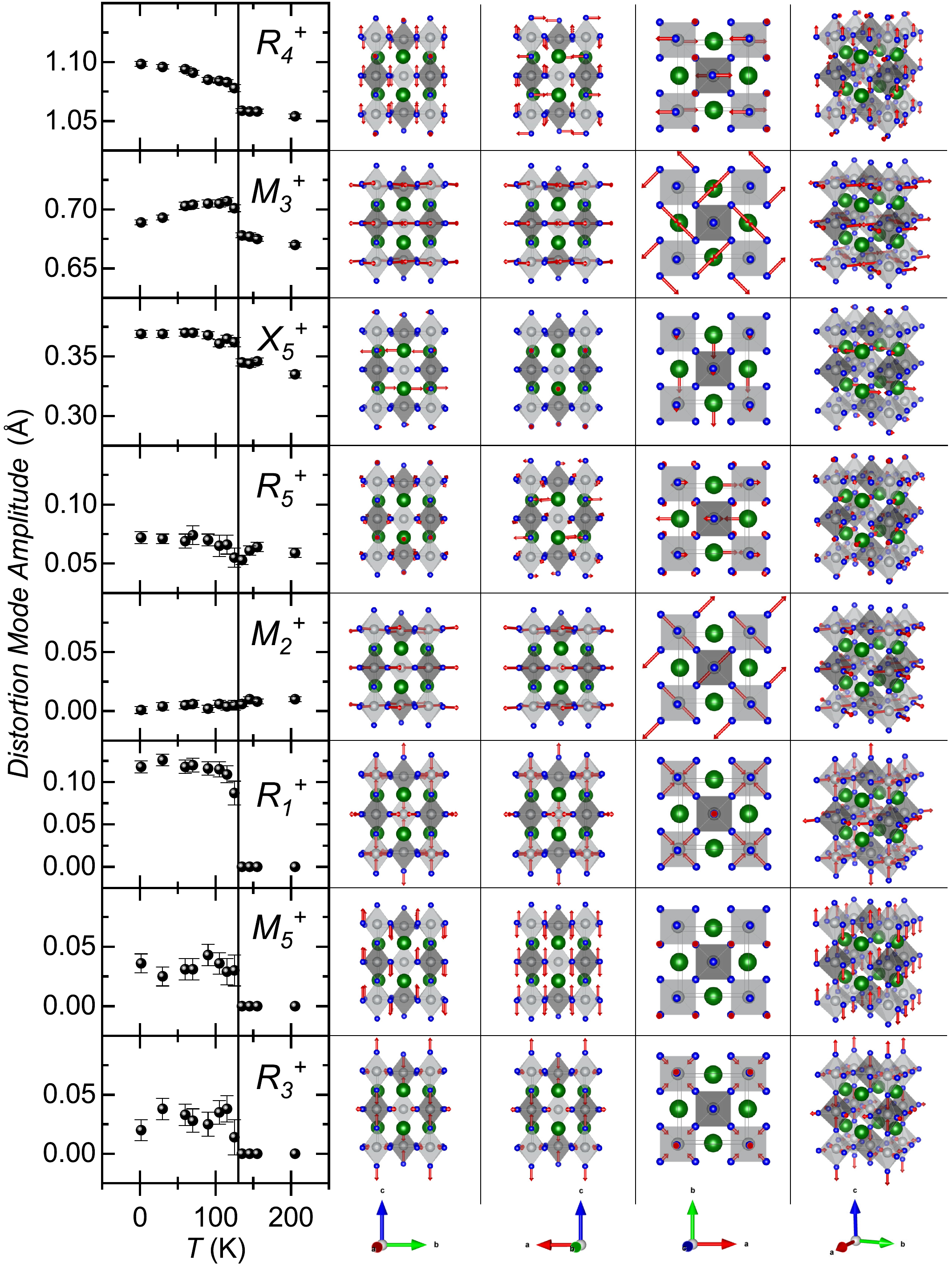}
	\vspace{-3mm}
	\caption{Temperature dependence of the PrNiO$_3$ (S2) global mode amplitudes across $T_{MIT}$ obtained from the fits of HRPT data, together with a schematic representation of the atomic displacements. For each $irrep$, the atoms involved in the distortion mode(s) with this symmetry are indicated. For the $irreps$ with more than one mode ($X^{+}_5$ and $R^{+}_5$), the arrow lengths reflect the refined relative amplitudes of the individual modes in the insulating $P2_1/n$ phase. To make comparisons easier, the scale in the vertical axis is the same for all mode amplitudes.}
	\label{fig:Global_mode_amplitudes}\vspace{-5mm}
\end{figure*}

\begin{figure*}[tbh!]
	\includegraphics[keepaspectratio=true,width=154 mm]{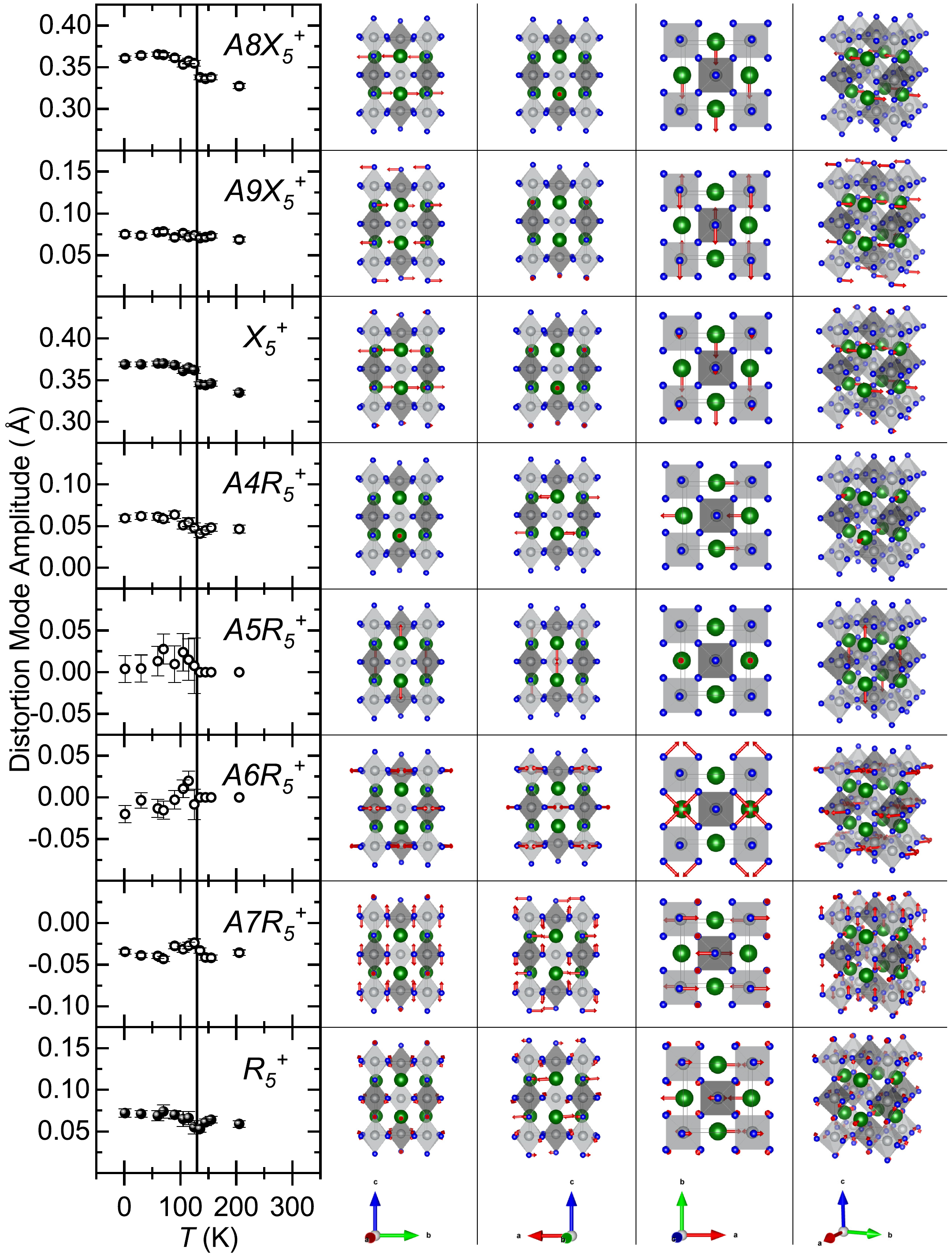}
	\vspace{-3mm}
	\caption{Temperature dependence of the PrNiO$_3$ (S2) individual (open symbols) and global (full symbols) mode amplitudes across the MIT for the modes of $X^{+}_5$ and $R^{+}_5$ symmetry, as obtained from HRPT data fits.  For each individual mode, the $AMPLIMODES$ label is indicated. In the case of the global amplitudes, the arrow lengths in the right panels reflect the refined relative amplitudes of the individual modes in the insulating $P2_1/n$ phase. To make comparison easier the scale in the vertical axis is the same for all mode amplitudes.}
	\label{fig:Individual_mode_amplitudes}\vspace{-5mm}
\end{figure*}

Schematic representations of the $P2_1/n$-allowed modes associated to the $R^{+}_1$, $M^{+}_5$ and  $R^{+}_3$ irreps are shown in Fig.~\ref{fig:Global_mode_amplitudes} together with the temperature dependence of their amplitudes. For the two extra modes with $R^{+}_5$ symmetry, schematic representations together with the temperature dependence of the individual and the global amplitudes are shown in Fig.~\ref{fig:Individual_mode_amplitudes}. A list of the basis vectors associated to each $irrep$ is provided in the SUPP2 file of the Supplemental Material.

\section{Temperature dependence of the mode amplitudes}

\subsection{Metallic phase}
As shown in Fig.~\ref{fig:Global_mode_amplitudes}, the largest amplitude values at any temperature are those of the $R^{+}_4$ ($\sim$ 1.06 \AA) and $M^{+}_3$ ($\sim$  0.67 \AA) oxygen modes. The prime contribution of these two distortion modes to the \textit{Pbnm} structure was predicted in the past from the rationalization of rigid octahedron tilting schemas in perovskites~\cite{Glazer1972,Woodward_1997a,Woodward_1997b}. In Glazer's terminology, it can be expressed as \textit{a$^-$b$^-$c$^+$} (pseudo-cubic notation), the positive/negative signs meaning that the rotations are in phase/out of phase for successive octahedra along the same axis. More recently, such distortion has been also interpreted in terms of Landau Theory as the result of the condensation of two order parameters with \emph{\textbf{k}} =(1/2, 1/2, 1/2) and \emph{\textbf{k}} = (1/2, 1/2, 0)~\cite{AMPLIMODES1,AMPLIMODES2}.

Application of symmetry arguments shows that distortion modes belonging to other irreps and involving not only oxygen sites, but also rare-earth sites, are also possible, and this is indeed confirmed by our analysis. As shown in  Fig.~\ref{fig:Individual_mode_amplitudes}, the amplitudes of the oxygen modes with $X^{+}_5$ (A9) and $R^{+}_5$ (A7) symmetry are very small for both irreps above $T_{MIT}$ ($\sim$ 0.07 \AA), but nonzero within the experimental error. The praseodymium displacements along \emph{\textbf{a}} in $R^{+}_5$ are also of this order of magnitude, but those along \emph{\textbf{b}} of  $X^{+}_5$  symmetry are significantly larger ($\sim$  0.3 \AA ). The $M^{+}_2$ mode amplitude is in contrast nearly zero ($\sim$ 0.01 \AA), at least within our experimental resolution. Tables with the amplitudes and their sigmas at all the investigated temperatures are provided in the SUPP3 file of the Supplemental Materials.

The temperature dependence of all mode amplitudes above $T_{MIT}$ is approximately linear, but two different slopes are clearly observed. As we will see later (see section "Reproducibility"), this behavior is confirmed by an independent data set measured in a different PrNiO$_3$ sample up to RT.  We note also that the global amplitudes of $R^{+}_4$ and $R^{+}_5$ are approximately constant above $T_{MIT}$. In contrast, those of $M^{+}_3$ and $X^{+}_5$ increase by cooling at a much faster rate of about $10^{-5}$ \AA/K. This behaviour suggests that each primary mode has an associate secondary (i.e., induced) mode: $R^{+}_5$ is that of $R^{+}_4$, and $X^{+}_5$  that of $M^{+}_3$. The coupling of these last two modes indicates that, in the metallic phase,  the main impact in the rare earth displacements along \emph{\textbf{b}} comes from the octahedral rotations around the \emph{\textbf{c}} axis ($M^{+}_3$), and not from those around \emph{\textbf{c}} ($R^{+}_4$). Moreover, the pronounced temperature dependence of both modes suggest that, in spite of having amplitudes much smaller than $R^{+}_4$, they are the main responsible of the evolution of the crystal structure between RT and $T_{MIT}$.

\begin{figure}[tbh!]
	\includegraphics[keepaspectratio=true,width=\columnwidth]{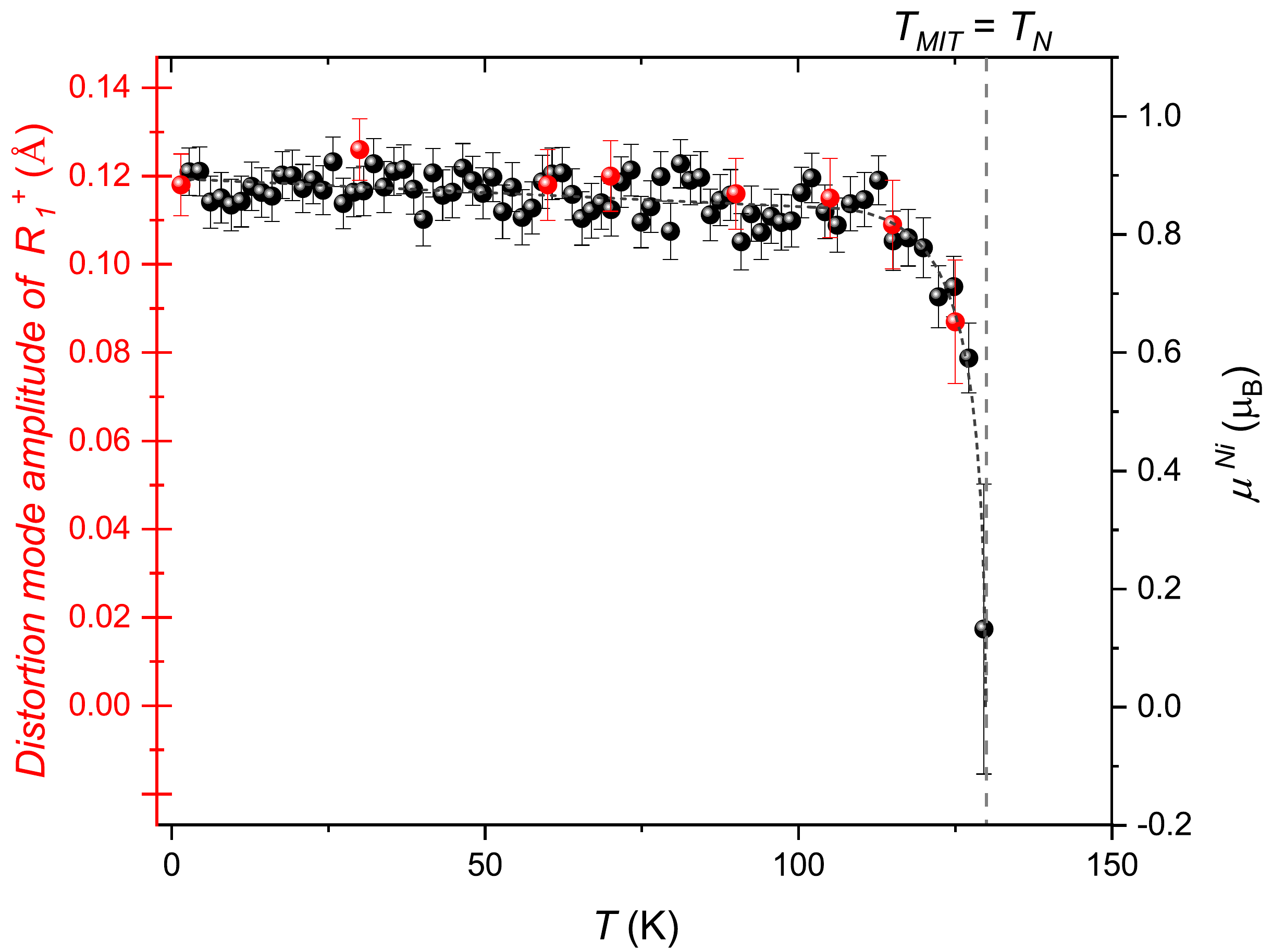}
	\vspace{-3mm}
	\caption{Temperature dependence of the $R^{+}_1$ mode amplitude as obtained from HRPT data refinements (left axis) and the Ni staggered magnetization as obtained from DMC data fits (right axis) for PrNiO$_3$ (S2). The vertical dashed line marks  $T_{MIT}$ =  $T_N$  ($\sim$ 130 K). The small-dashed line is a guideline for the eye.}
	\label{fig:Order_parameters}\vspace{-5mm}
\end{figure}

\subsection{Insulating phase}
The approximately linear temperature dependence of the global mode amplitudes observed in the metallic phase is abruptly interrupted at the MIT. Below the transition $all$ modes of $Pbnm$ parentage undergo sharp anomalies involving amplitude increases between 0.02 and 0.15 \AA. The only exception is $M^{+}_2$, whose amplitude remains close to zero also in the insulating phase. Interestingly, their temperature dependencies display some important differences with respect to the behavior observed in the metallic state.

As shown in Fig.~\ref{fig:Global_mode_amplitudes}, none of the five global amplitudes present a simple linear behavior below $T_\mathrm{MIT}$. Moreover, only those belonging to $R^{+}_4$, $X^{+}_5$ and $R^{+}_5$ keep increasing by decreasing temperature. In contrast, the $M^{+}_3$ amplitude decreases in the insulating phase after the jump at $T_\mathrm{MIT}$. This observation suggests that the in-plane rotation around  \emph{\textbf{c}} is the only $Pbnm$ mode destabilized by the appearance of the new monoclinic modes.  A further observation is the presence of a small anomaly around 60 K, clearly visible in the two modes with the largest amplitudes ($R^{+}_4$ and $M^{+}_3$).

The evolution of the individual $X^{+}_5$ and $R^{+}_5$ mode amplitudes in the insulating phase is shown in Fig~\ref{fig:Individual_mode_amplitudes}. For  $X^{+}_5$, the amplitude of the oxygen displacements along  \emph{\textbf{b}} (A9) remains small without any significant change across the MIT. In contrast, the praseodymium displacements along this direction (A8) increase substantially, being the main contributors to the jump in the global amplitude. In the case of $R^{+}_5$,  the amplitude of the oxygen displacements of $Bmab$ parentage  (A7) remains also small and nearly constant. The praseodymium displacements along \emph{\textbf{a}} (A4), in contrast, undergo a small increase.  The two additional monoclinic Pr (A5) and O (A6) modes of $R^{+}_5$ symmetry, which are forbidden in the metallic phase, display zero amplitude also below the MIT within our experimental resolution.

We discuss now the modes of the new $irreps$ $M^{+}_5$, $R^{+}_3$, and $R^{+}_1$ absent in the orthorhombic phase.  As shown in Fig~\ref{fig:Individual_mode_amplitudes}, the amplitude of $M^{+}_5$ and $R^{+}_3$ is rather small but non-zero below $T_{MIT}$. The breathing mode $R^{+}_1$, in contrast undergoes a huge jump of $\sim$ 0.12 \AA, much larger than the anomalies observed in the remaining modes, and in particular the Jahn-Teller distortion ($R^{+}_3$). This clearly signals the breathing mode as the leading structural response associated to the electronic instability at $T_{MIT}$, at least for PrNiO$_3$. However, it does not exclude that, as suggested by some authors~\cite{Mazin2007,Peil2019} the relative amplitudes of the breathing and JT modes could change along the series, i.e., by moving towards a more insulating ground state.

A further interesting observation is that the temperature dependence of the breathing distortion amplitude follows closely that of the staggered magnetization (Fig.~\ref{fig:Order_parameters}). This is clearly not the case for the two largest modes $R^{+}_4$ and $M^{+}_3$, whose pronounced temperature dependence below $T_{MIT}$ strongly differs from the nearly temperature-independent amplitude of the breathing mode and the staggered magnetization. This observation implies a linear coupling of both order parameters, further supporting a scenario where the magnetic order just follows the electronically-driven stabilization of the breathing mode.

 It is also worth stressing that the emergence of the breathing mode has important consequences for the pre-existing orthorhombic distortions, which become more pronounced (and hence more stable) in the insulating phase. The only exception is the in-plane rotation $M^{+}_3$, whose amplitude decrease indicates a progressive destabilization with the development of the breathing distortion. This suggests that the modeling of the changes in the electronic structure across the MIT should explicitly consider the evolution of all modes across the MIT, and not only that of the breathing mode. Moreover, it further confirms the high responsiveness of $M^{+}_3$, which reacts in a particularly pronounced (and distinct!) way to the electronic, magnetic, and structural changes at the transition. It also suggests that the role of this mode in the MIT could be more important than foreseen.

\begin{figure}[tbh!]
\includegraphics[keepaspectratio=true,width=\columnwidth]{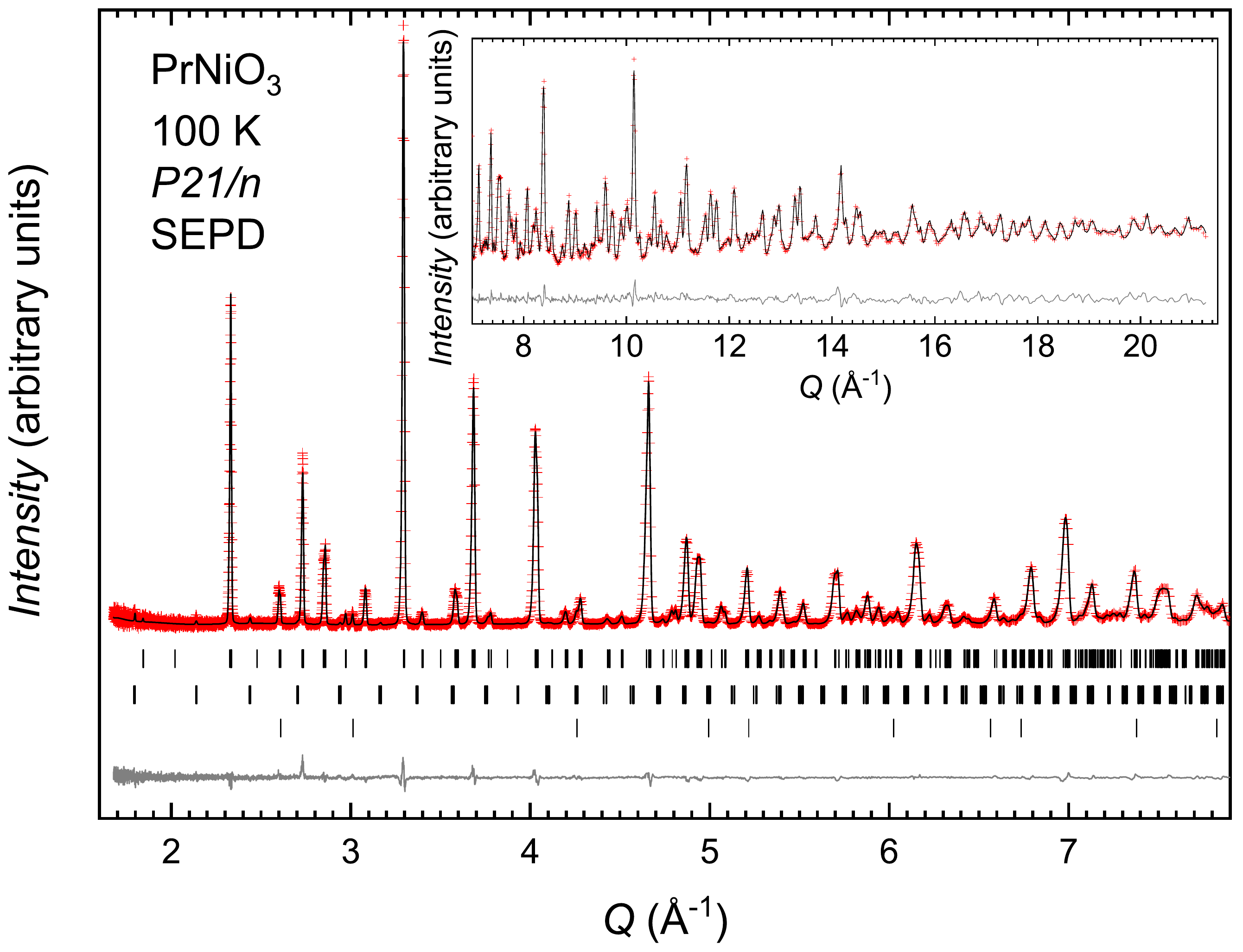}
\vspace{-3mm}
\caption{Rietveld fit of the SEPD powder neutron diffraction data for PrNiO$_3$ (S1) measured at 100 K. Red crosses: observed data. Black lines: calculated patterns. The vertical ticks indicate the positions of the Bragg reflections for the crystal structure of PrNiO$_3$ (upper row) and NiO impurities (middle row), and the collinear antiferromagnetic magnetic structure of PrNiO$_3$ with \emph{\textbf{k}} = (1/2, 0 ,1/2) (lower row) described in the text. }
\label{fig:SEPD_fits}
\end{figure}

\begin{figure}[tbh!]
\includegraphics[keepaspectratio=true,width=\columnwidth]{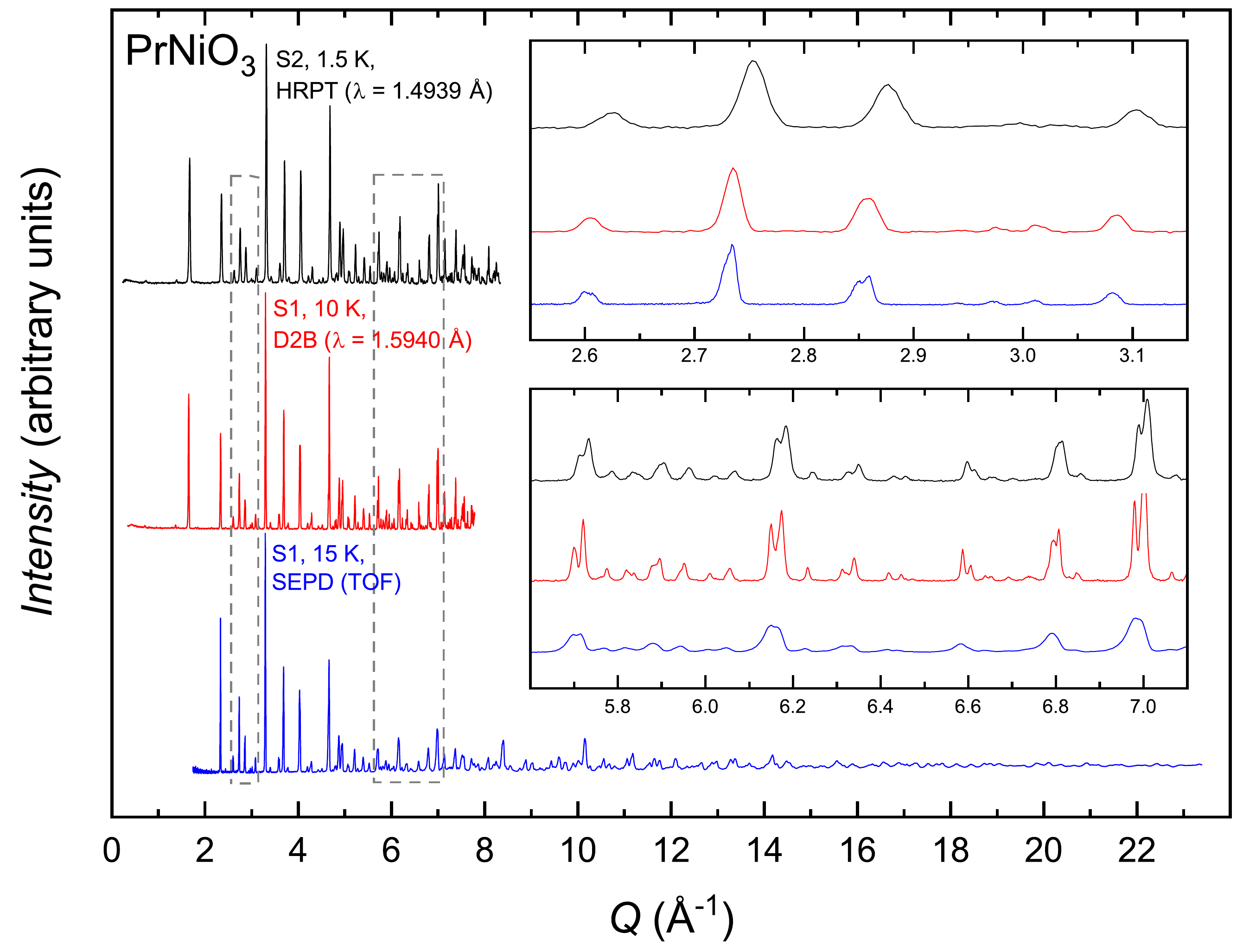}
\vspace{-3mm}
\caption{Comparison between the data recorded on HRPT (S2), D2B (S1) and SEPD (S1) for PrNiO$_3$. Inset: magnification of selected Q-regions.}
\label{fig:Comparison_instruments}\vspace{-5mm}
\end{figure}

\subsection{Reproducibility}

The distortion mode analysis of the HRPT neutron powder diffraction data presented in previous sections involves an important number of new results. Besides providing the first experimental determination of the different mode amplitudes in a large temperature region covering both sides of the MIT, they also point out  to important aspects previously unnoticed or even ignored. It is thus important to check the robustness of the methodology, in particular against different data sets. With this idea in mind we have  used the distortion mode formalism to re-analyze two older sets of temperature dependent neutron powder diffraction data recorded on different instruments. The same large PrNiO$_3$ powder sample (about 5g, referred here as S1) was used.

The first data set, obtained on the high resolution powder diffractometer D2B at the Institute Laue Langevin in Grenoble, France, was the same used in reference~\cite{Medarde2008}. The data were recorded using the high resolution mode, with Ge[335] $\lambda$ = 1.594 \AA \, (\textit{Q}$_{max}$ = 7.77 {\AA}$^{-1}$). The second set (unpublished) was obtained at time-of-flight medium resolution diffractometer SEPD at Argonne National Laboratory, USA (\textit{Q}$_{max}$ = 21.27 {\AA}$^{-1}$). In both instruments several patterns were recorded as a function of temperature (10 - 170 K for D2B,  10 K - RT for SEPD). All patterns were measured by heating after a stabilization time of about 30 minutes with typical acquisition times 5 - 6 hours per temperature. A representative pattern showing the refinement of the data obtained in SEPD at $T$ = 100 K is shown in Fig.~\ref{fig:SEPD_fits}. A comparison of the data obtained on the three instruments  is shown in Fig.~\ref{fig:Comparison_instruments}.

\begin{figure}[tbh!]
\includegraphics[keepaspectratio=true,width=\columnwidth]{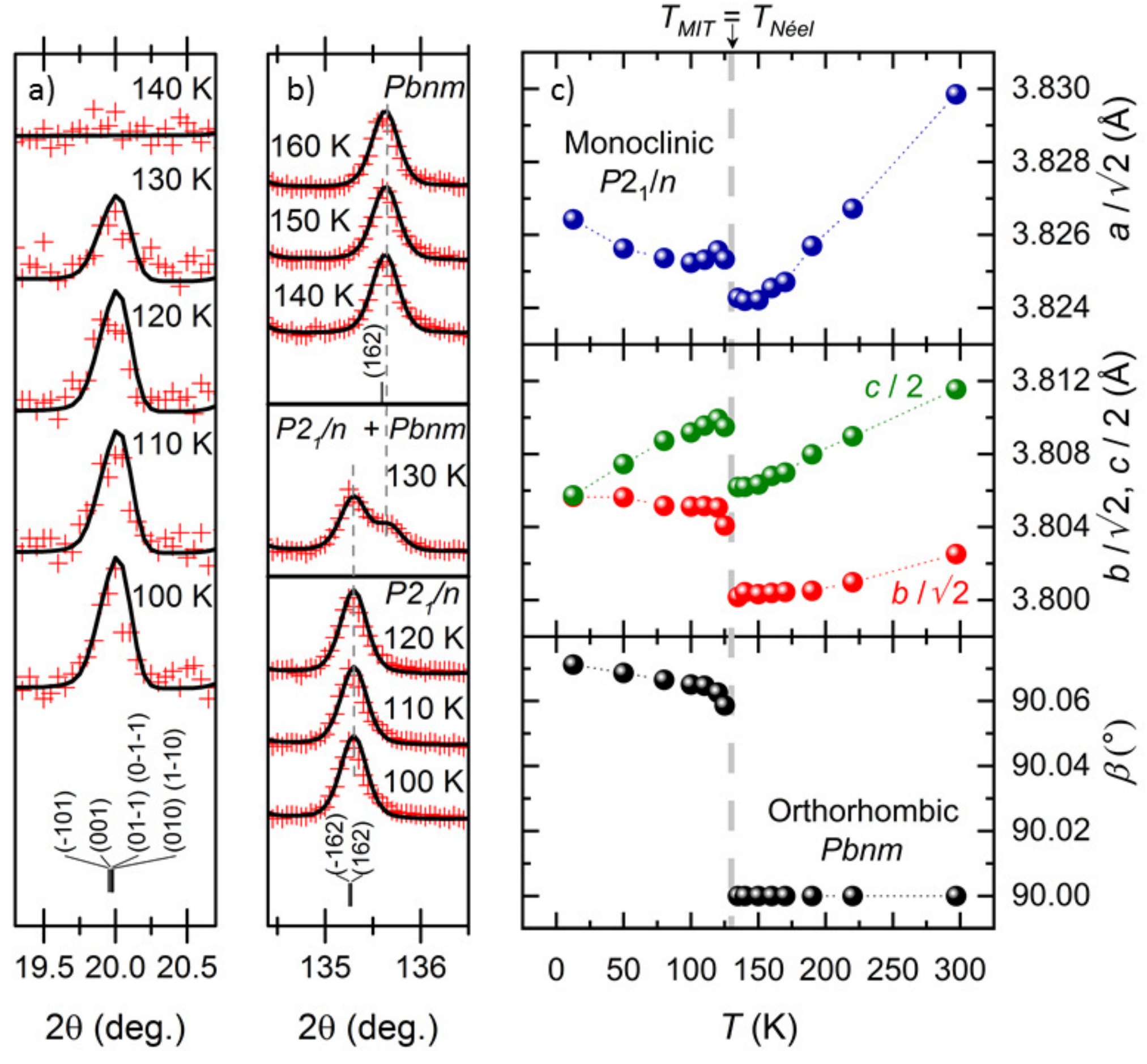}
\vspace{-3mm}
\caption{(a) Temperature dependence of  the magnetic reflection (1/2, 0, 1/2) (a) and the nuclear reflection (162) (b) of PrNiO$_3$ (S1) across $T_{MIT}$. A coexistence of the $Pbnm$ and $P2_1/n$ phases is only observed at $T_{MIT}$ = $T_N$ = 130 K. (c) Temperature dependence of the lattice parameters as obtained from D2B and SEPD fits. Error bars are smaller than data points indicators.}
\label{fig:Lattice_D2B_SEPD}
\end{figure}

In spite of the different origin and slightly different preparation methods, the old PrNiO$_3$ sample (S1) was found to be very similar to the one described in previous sections (S2) in terms of lattice parameters, phase coexistence and metal-to-insulator temperature.   Figs.~\ref{fig:Lattice_D2B_SEPD}a and b show the temperature dependence of the magnetic reflection (1/2 0 1/2) and the nuclear reflection (162) across $T_{MIT}$, as measured on D2B by heating. In the low temperature monoclinic phase the doublet (-116)+(116) and the magnetic reflection (1/2 0 1/2) are clearly visible. At $T_{MIT}$ =  $T_N$ = 130 K the $Pbnm$ reflection (162) is also observed, indicating a coexistence of the $Pbnm$ and $P2_1/n$ phases at this temperature. For $T \, \geq$ 140 K the reflections corresponding to the monoclinic and the magnetic phases are absent, and only the $Pbnm$ phase subsists. These observations indicate that the sample is single phase at all the temperatures investigated with exception of $T_{MIT}$ = $T_N$ = 130 K. The macroscopic phase coexistence region is thus  $T_{MIT}$ $\pm$ 10 K, nearly identical as in the sample measured on HRPT.

The evolution of the pseudocubic lattice parameters and the monoclinic angle $\beta$ between room temperature (RT) and 10 K obtained from the joint fits of the D2B and SEPD data is shown in Fig.~\ref{fig:Lattice_D2B_SEPD}a. \footnote{
The $a$, $b$ and $c$ values at 50 and 80 K were obtained from fits of SEPD data alone. Due to the superior resolution of D2B, the monoclinic angle at these two temperatures was fixed to the values obtained with combined SEPD+D2B fits}. The sharp anomalies displayed by $a_p = a/ \surd2, \, b_p = b/\surd2$, and $c_p = c$/2 at $T_{MIT}$ are consistent with the narrow phase coexistence region around the transition. The anomaly around 60 K is also clearly observed.

For the Rietveld refinements of the D2B and SEPD data we used the same distortion mode decomposition described in previous sections. In a first series of fits we fixed to zero the amplitude of the modes which, from the analysis of the HRPT data, were found to display negligibly small amplitudes within the experimental standard deviations in the full temperature range. This concerns only the $R^{+}_5$ Pr (A5) and O (A6) modes of \textit{P2$_1$/n} parentage (note that they are symmetry forbidden in the metallic phase). The amplitude of the $M^{+}_2$ mode, allowed in both space groups and with a much smaller standard deviation, was in contrast refined in the full temperature range.

\begin{figure}[tbh!]
\includegraphics[keepaspectratio=true,width=\columnwidth]{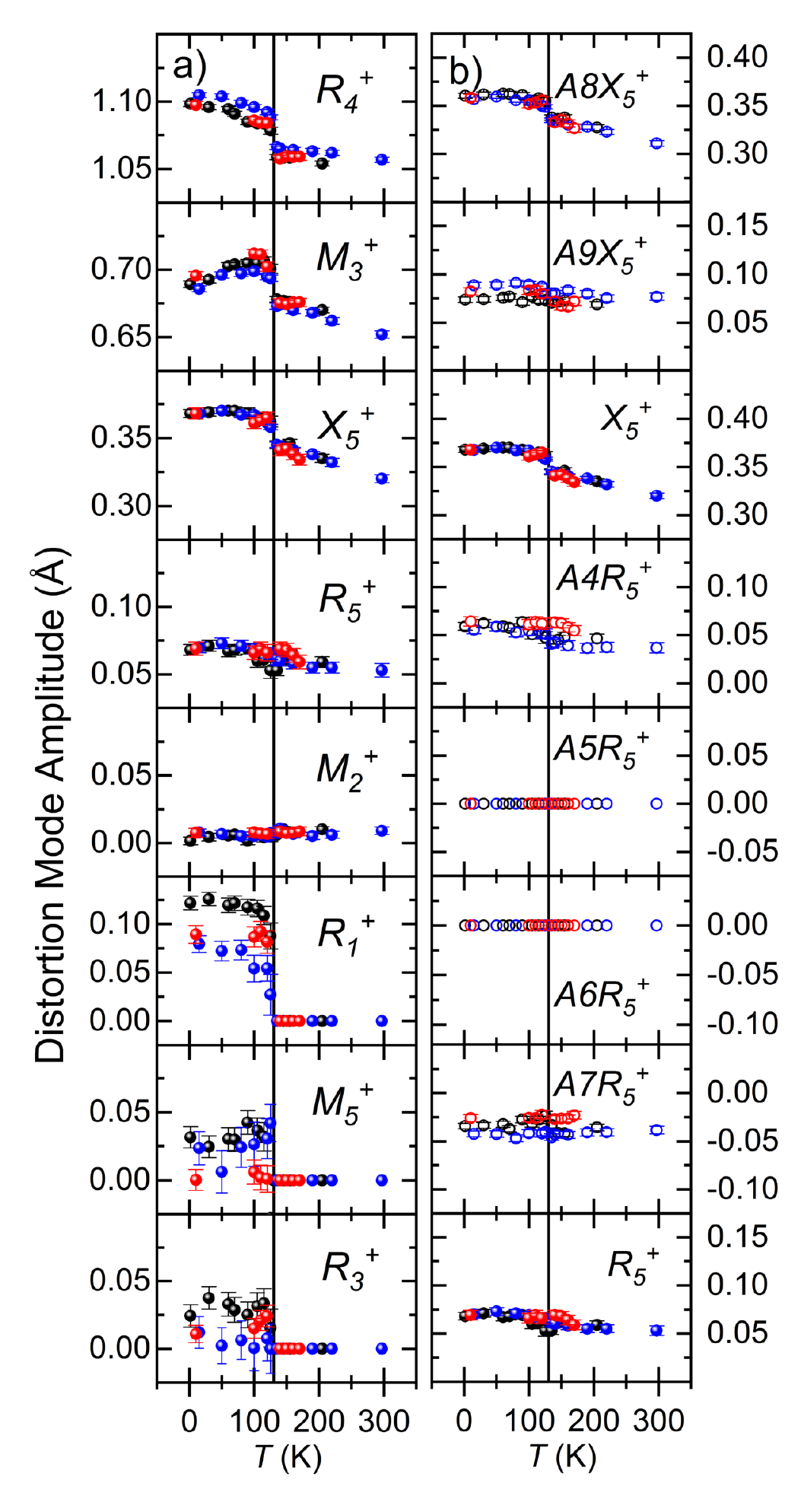}
\vspace{-3mm}
\caption{Temperature dependence of the PrNiO$_3$ mode amplitudes obtained with HRPT data (S2, black points) with those obtained from D2B (S1, red points) and SEPD (S1, blue points). The data were obtained after nullifying the amplitudes of the purely monoclinic Pr (A5) and O (A6) modes of $R^{+}_5$ symmetry below $T_{MIT}$. (a) Global amplitudes. (b)  Individual mode amplitudes of $X^{+}_5$ and $R^{+}_5$ symmetry together with their associated global amplitudes.}
\label{fig:Comparison_global_amplitudes}\vspace{-5mm}
\end{figure}

The obtained results for the global and individual mode amplitudes are shown in Fig.~\ref{fig:Comparison_global_amplitudes} together with those obtained from HRPT data. Altogether we found a very good agreement between the three sets of data, indicating a very good reproducibility when the described fitting strategy is used. This is remarkable given the large differences in resolution and $Q$-range of the three instruments, which may result in different estimations of the background and the Debye-Waller factors, and hence, of the integrated intensities. The small discrepancies in the global amplitudes of the  $R^{+}_1$,  $R^{+}_3$ and  $R^{+}_4$ modes are probably due to such differences. In the particular case of the breathing mode amplitude, we note that SEPD fits tends to provide lower values than high resolution instruments such as D2B and HRPT, in particular close to the transition where the monoclinic angle is smaller and the degree of superposition in the powder diffraction patterns higher.

\begin{figure}[tbh!]
\includegraphics[keepaspectratio=true,width=\columnwidth]{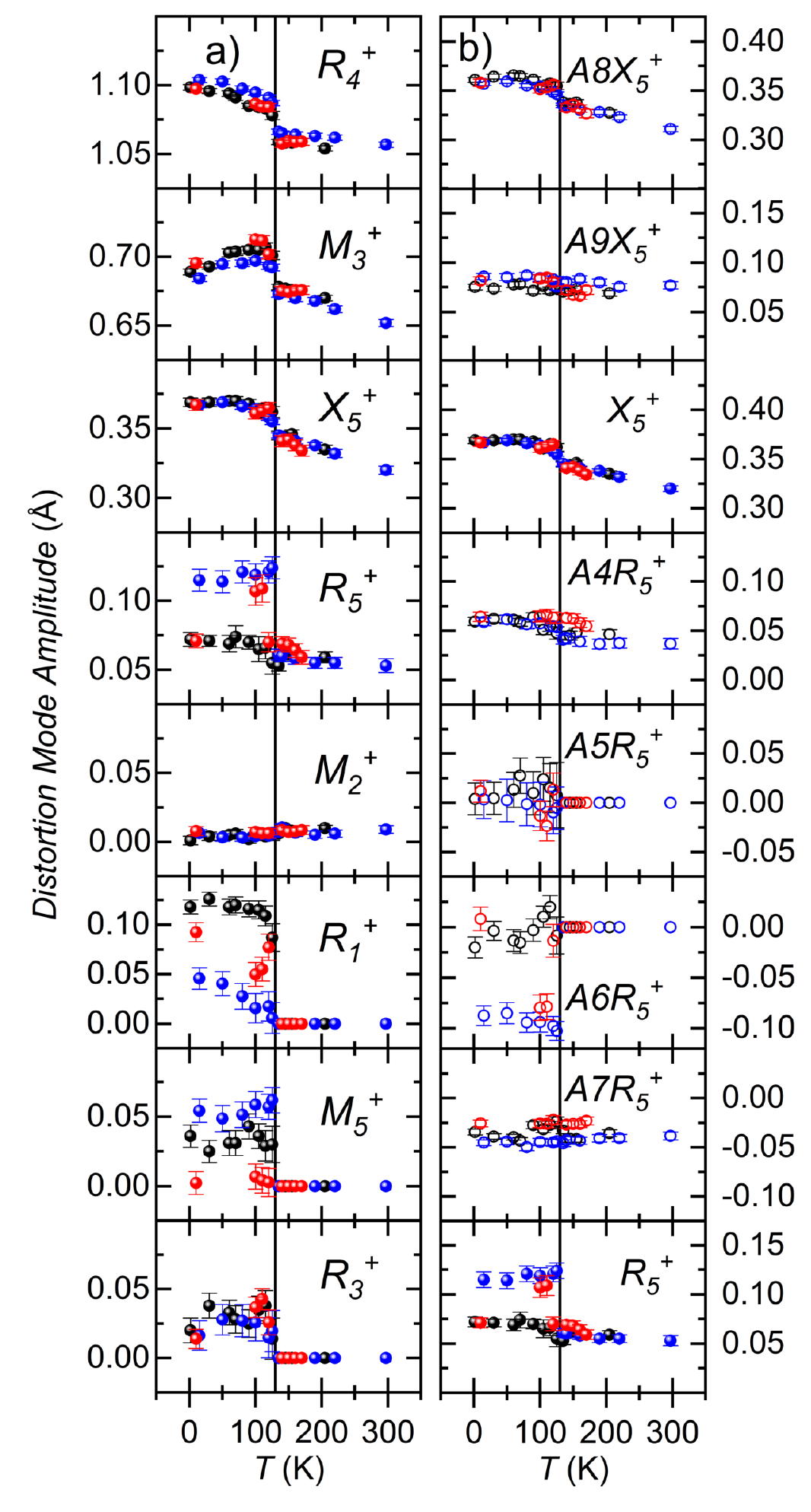}
\vspace{-3mm}
\caption{Temperature dependence of the PrNiO$_3$ mode amplitudes obtained with HRPT data (S2, black points) with those obtained from D2B (S1, red points) and SEPD (S1, blue points). The data were obtained after refining the amplitudes of all the new monoclinic modes below $T_{MIT}$. (a) Global amplitudes. (b)  Individual mode amplitudes of $X^{+}_5$ and $R^{+}_5$ symmetry together with their associated global amplitudes.}
\label{fig:Comparison_individual_amplitudes}\vspace{-5mm}
\end{figure}

In a second series of fits we allowed all mode amplitudes to be refined. Although most of them remained unchanged, the global amplitudes obtained from SEPD fits in the insulating phase were significantly different for two $irreps$:  $R^{+}_5$, with four contributing modes (A4, A5, A6 and A7) below $T_{MIT}$, and $R^{+}_1$, with a single one (the breathing mode), see Fig.~\ref{fig:Comparison_individual_amplitudes}. Moreover, both global amplitudes displayed evident signs of correlation. A similar result was obtained in joint fits combining data from D2B (the instrument with the highest resolution of this study) with those of SEPD. Looking to the individual amplitudes, also shown in Fig.~\ref{fig:Comparison_individual_amplitudes}, it is straightforward to see that the origin of the behavior is the anomalously large amplitude of one of the two new modes $R^{+}_5$ symmetry that appear in the insulating phase (A6, involving displacement of the basal oxygens). Given that both, HRPT and D2B fits give nearly zero amplitude for this mode at low temperatures, where breathing distortion is well established, the SEPD results are probably due to the impossibility to resolve such a small structural details with a medium-low resolution instrument.

These results stress the importance of a) choosing an instrument adapted to the complexity of the structure for precise structural studies and b) making reasonable choices for the restrictions applied. In the particular case of PrNiO$_3$, the limiting parameter appears to be the level of superposition of the Bragg reflections in the monoclinic phase, particularly high in the vicinity of $T_{MIT}$ where the monoclinic angle is small. This favors high resolution machines with moderate $Q$-range such as D2B and HRPT over medium/low resolution machines covering a large $Q$-range. An important point is the fact that the results obtained on SEPD can also provide a good, correlation-free description of the structure if the fits are carried out with the proper constraints. In this case, the criterion is given by the values of the experimental sigmas, which indicate which amplitudes may be susceptible to be nullified. Systematic comparison of the results obtained with different instruments appear in any case as desirable to provide definitive answers concerning the actual values of small mode amplitudes.

\section{Anomalies below $T_\mathrm{MIT}$ }

Besides the lattice, electric and magnetic anomalies at the MIT, we showed in previous sections the presence of a second discontinuity around $\sim 0.4 \times\ T_\mathrm{MIT} \sim$ 55 - 60 K. This anomaly, clearly visible in the electric resistivity, the lattice and other structural parameters, marks the lower limit of a region extending up to $T_{MIT}$ where the behavior of these properties is anomalous. We also showed that an origin based on the coexistence of the metallic and insulating phases is neither supported by resistivity nor by XRD data. Indeed, these techniques suggest that the nucleation of the metallic phase starts around 90 K in the form of microscopic domains. Such domains would growth in size by increasing temperature, reaching sizes large enough to be observed by XRD (4 - 5 nm) around 116 K.

In spite of these observations, the anomalous behavior of the lattice below 90 K suggest that, at least by heating, the structural and electronic reorganization below $T_{MIT}$ starts at significantly lower temperatures, around 55 - 60 K. This unusual behavior displays a close resemblance with the temperature dependence of the DOS in the vicinity of $E_F$ reported in ref.~\cite{Vobornik1999} for bulk PrNiO$_3$, NdNiO$_3$ and other solid solutions with $T_{MIT}$ = $T_N$. As shown by Vobornik and co-workers, the DOS is not fully suppressed at $T_{MIT}$ in none of these materials. Instead, it smoothly decreases down to a normalized temperature $T^* \sim 0.4 \times\ T_\mathrm{MIT}$ common to all the samples investigated. This contrasts with the behavior of the nicklelates with $T_{MIT}$ $>$ $T_N$ (R = Sm and Eu), where the DOS is immediately suppressed a few degrees below $T_{MIT}$. To explain these observations these authors suggested the existence of a possible interplay between electronic and magnetic degrees of freedom in the nicklelates where the two transition temperatures coincide. Even if the electronic anomaly seems to be the leading mechanism, that would imply that the gap opens "smoothly" when the magnetic order is already well established, as in the case of PrNiO$_3$ (see Fig.~\ref{fig:Physical_properties}f). Interestingly, the value of  $T_{MIT} \times\  0.4$ for our PrNiO$_3$ sample is 52 K, very close to the anomaly observed in the lattice parameters, interatomic distances and mode amplitudes, suggesting a possible connection between Vobornik's observations and our data.

A possible mechanism able to provide an interplay between electronic, magnetic and lattice degrees of freedom below $T_{MIT}$ = $T_N$ could be the presence of magnetism-induced polar distortions.  Such scenario has been theoretically predicted,~\cite{Perez-Mato2016,Giovannetti2009} and the value of the spontaneous electric polarization in the insulating antiferromagnetic phase estimated for several nickelates using ab-initio calculations. This prediction could not be verified to date due to the leaky nature of the early nickelates -the only available as thin films-, as well as to the absence of single crystals for the late (probably more insulating) members of the $R$NiO$_3$ family. However, it is worth noting that if this prediction turns out to be true, the actual symmetry of the crystal structure below $T_\mathrm{N}$ should be lower due to the loss of the inversion center, and this will allow the existence a larger number of symmetry-adapted distortion modes below $T_\mathrm{N}$. That would increase the number of candidates to display nonzero amplitude in the insulating phase competing with the pre-existing orthorhombic distortions. Within this scenario the structural anomalies at 55-60K would be the result of a change in the relative stability of the different modes, which could in turn result in a change of the Ni 3d - O 2p orbital overlap and hence in an anomaly in the electric resistivity. Experiments on the late nickelates, where  $T_\mathrm{MIT}$ and $T_\mathrm{N}$ are well separated, will be highly desirable to asess the validity of this scenario.

\section{Summary and conclusions}

To summarize, we have presented a detailed study of bulk PrNiO$_3$ which combines physical properties and accurate structural data covering a large temperature window across $T_{MIT}$. We have compared the thermal behavior of the electric resistivity, magnetization, and heat capacity of PrNiO$_3$ with that of high resolution neutron and synchrotron x-ray powder diffraction data that, in contrast to previous studies, we analyzed in terms of symmetry-adapted distortion modes. This analysis allows to determine the evolution of the individual mode amplitudes across the transition. Such amplitudes can be seen as order parameters for the different individual distortions. Hence, their relative values, temperature dependence and anomalies at $T_{MIT}$ encode the information on the most relevant degrees of freedom involved in the global distortion and their evolution across the MIT.

Using this formalism, we have shown that the anomalies at the MIT, traditionally described in terms of changes in the interatomic distances and angles, appear as abrupt increases of the orthorhombic mode amplitudes accompanied by the emergence of new modes below the transition. Our data also allowed to classify them according to their magnitude and temperature dependence, and to establish primary-secondary mode coupling schemas. They also allowed to identify the in-plane rotation $M^{+}_3$ as the main responsible of the evolution of the crystal structure in the metallic phase.

The largest anomaly at $T_{MIT}$ was observed in the amplitude of the breathing mode $R^{+}_1$, which undergoes a sharp jump of 0.15 \AA. This clearly signals $R^{+}_1$ as the leading structural response associated to the electronic instability at $T_{MIT}$, at least for PrNiO$_3$. We have also shown that the temperature dependence of the $R^{+}_1$ amplitude follows closely that of the staggered magnetization. This observation implies a linear coupling of both order parameters, further supporting a scenario where the magnetic order just follows the electronically-driven stabilization of the breathing mode.

Since this kind of analysis is reported for the first time, we dedicated a section to show that its results are compatible with those of previous studies, in particular in what concerns the evolution of the interatomic distances and angles across the MIT. We also showed that the obtained mode amplitudes are fully reproducible through a comparison of the results obtained from the analysis of two additional data sets on a different sample measured in two neutron powder diffractometers with different resolutions and $Q$-ranges. Such comparison showed the importance of choosing an instrument adapted to the complexity of the structure for precise structural studies, and pointed out the need of making reasonable restrictions when low resolution data are used.

To conclude, we would like to mention that our data also uncover an unexpected, previously unnoticed transient regime characterized by anomalous temperature dependence of the lattice parameters and several distortion mode amplitudes that persists down to $T^*$  $\sim$ 0.4 x $T_{MIT}$.  Since phase coexistence is only observed in a small temperature region around $T_\mathrm{MIT}$, these observations suggest the existence of a competition between the high and low-temperature modes that persists well below $T_\mathrm{MIT}$. Given that anomalous behavior of the DOS around $E_F$ has been reported in the same temperature interval for the nickelates with $T_\mathrm{MIT}$ = $T_\mathrm{N}$, our observations may be specific of the $RNiO_3$ compounds where the MIT and the transition to the N\'eel state coincide. If this is the case, the recently predicted existence of magnetism-driven polar distortions below $T_{N}$ could be at the origin of this intriguing behavior.  Although a definitive answer can only be provided by temperature-dependent diffraction studies for the nickelates with $T_{MIT}$ $\neq$ $T_{N}$, any future model of the MIT in $R$NiO$_3$ will have to address these novel experimental facts.

\section{Acknowledgements}

This work has benefited from stimulating discussions with C. Ederer, A. Hampel, O. Peil and A. Georges. It is based on experiments performed at the Swiss spallation neutron source SINQ, Paul Scherrer Institute (Villigen, Switzerland), the Institut Laue Langevin, (Grenoble, France) and IPNS (Argonne, USA). We thank the Swiss National Center of Competence in Research MARVEL (Computational Design and Discovery of Novel Materials) for financial support, as well as the allocation of  beam time by the PSI, ILL and IPNS.

\bibliographystyle{apsrev}
\vspace{-7mm}
\bibliography{BIB_RNiO3}

\end{document}